\documentclass[review]{elsarticle}

\usepackage{geometry}
\usepackage[colorlinks=green,
linkcolor=red,
anchorcolor=blue,
citecolor=green]
{hyperref}

\usepackage{times}
\usepackage{soul}
\usepackage{url}

\usepackage[utf8]{inputenc}
\usepackage[small]{caption}
\usepackage{graphicx}
\usepackage{amsmath}
\usepackage{booktabs}
\usepackage{algorithmic}
\usepackage[ruled]{algorithm2e}
\usepackage{graphicx}  
\usepackage{subfigure}
\usepackage{multirow}
\usepackage{amsmath}
\usepackage{amsfonts}
\usepackage{color}
\usepackage{framed}
\usepackage{bbding}

\emergencystretch=\maxdimen
\journal{Journal of Computers \& Security}

\begin{document}

\begin{frontmatter}

\title{ActGraph: Prioritization of Test Cases Based on Deep Neural Network Activation Graph}


\author[mymainaddress,mysecondaryaddress]{Jinyin~Chen}
\ead{chenjinyin@zjut.edu.cn}

\author[mymainaddress]{Jie~Ge}
\ead{2112103116@zjut.edu.com}
\author[mymainaddress]{Haibin~Zheng\corref{mycorrespondingauthor}}
\ead{haibinzheng320@gmail.com}

\cortext[mycorrespondingauthor]{Corresponding author}

\address[mymainaddress]{College of Information Engineering, Zhejiang University of Technology, Hangzhou, 310023, China}
\address[mysecondaryaddress]{Institute of Cyberspace Security, Zhejiang University of Technology, Hangzhou, 310023, China}

\begin{abstract}
Widespread applications of deep neural networks (DNNs) benefit from DNN testing to guarantee their quality.
In the DNN testing, numerous test cases are fed into the model to explore potential vulnerabilities, but they require expensive manual cost to check the label.
Therefore, test case prioritization is proposed to solve the problem of labeling cost,
e.g.,
activation-based and
mutation-based prioritization methods.
However, most of them suffer from limited scenarios (i.e. high confidence adversarial or false positive cases) and high time complexity.
To address these challenges,
we propose the concept of the activation graph from the perspective of the spatial relationship of neurons.
We observe that the activation graph of cases that triggers the model's misbehavior significantly differs from that of normal cases.
Motivated by it, we design a test case prioritization method based on the activation graph,
ActGraph,
by extracting the high-order node features of the activation graph for prioritization.
ActGraph explains the difference between the test cases to solve the problem of scenario limitation.
Without mutation operations, ActGraph is easy to implement, leading to lower time complexity.
Extensive experiments on three datasets and four models demonstrate that ActGraph has the following key characteristics.
\textit{(\romannumeral1)}~\emph{Effectiveness and generalizability}: ActGraph shows competitive performance in all of the natural, adversarial and mixed scenarios, especially in \textit{RAUC-100} improvement ($\sim\times$1.40).
\textit{(\romannumeral2)}~\emph{Efficiency}:
ActGraph does not use complex mutation operations and runs in less time ($\sim\times$1/50) than the state-of-the-art method.


\end{abstract}

\end{frontmatter}

\section{INTRODUCTION\label{Introduction}}
Deep neural networks (DNNs) are widely applied in object recognition~\cite{zhang2022misleading}, image classification~\cite{duan2021mask}, natural language processing~\cite{alhogail2021applying}, autonomous vehicles~\cite{kim2021cybersecurity}, etc. But it is still threatened by uncertain inputs. For example, a car in autopilot mode recognized a white truck as a cloud in the sky and caused a serious traffic accident~\cite{2016tesla}. Therefore, it is crucial to test the DNN to find vulnerabilities before deployment.

In the testing phase, numerous diverse test cases are fed into the DNN in order to evaluate the reliability of the model.
These test cases require expensive manual labeling and verification.
To reduce the cost of labeling, a feasible solution is to prioritize test cases to find more vulnerabilities. These test cases are more likely to expose DNN vulnerabilities are marked in advance, improving the efficiency of DNN testing.

Some test case prioritization methods have been proposed to address the labeling cost problem. They can be divided into three categories, including Neuron Coverage (NC)-based~\cite{2017DeepXplore, 2018DLFuzz, 2018TensorFuzz, 2017DeepTest, 2018DeepGauge}, model activation-based~\cite{2020DeepGini, 2020Multiple, 2019Guiding, 2019Input} and mutation-based~\cite{2021Prioritizing} test case prioritization methods.
NC-based methods draw on the concept of test coverage~\cite{1997SoftwareZhu, 2002Software, 2009KLEE} from the traditional software testing to measure the adequacy of test cases.
But some studies~\cite{2019Structural, 2020Is, 2019There} have pointed out that it isn't strongly correlated between NC and misclassified inputs, so NC cannot be effectively applied to prioritization.

Model activation-based prioritization methods can be divided into confidence-based and embedding-based test case prioritization methods.
Confidence-based methods~\cite{2020DeepGini, 2020Multiple} extract the probability distribution of test cases in the DNN confidence layer. They believe that the correctly classified cases should output higher probabilities, but misclassified cases should output multiple similar probabilities. This assumption limits their application scenarios. For example, Carlini-Wagner (C\&W)~\cite{2017CW} adversarial cases can cause the DNN to output high confidence in the wrong class, so the effect of confidence-based methods will be significantly reduced.
On the other hand, embedding-based prioritization methods~\cite{2019Guiding, 2019Input} generally extract the hidden layer outputs of test cases.
They prioritize test cases with inconsistencies based on the differences between adversarial and normal cases.
However, they cannot effectively prioritize false positive (FP) cases, because the confusion of hidden layer features directly leads to the misclassification of the model.

Mutation-based test case prioritization, i.e., PRioritizing test inputs via Intelligent Mutation Analysis (PRIMA), is the state-of-the-art (SOTA) method, which uses mutation operations to make test cases as different as possible to produce different prediction results, arguing that it is easier to reveal the DNN vulnerabilities.
The time complexity of PRIMA is $O\left ( nm_{1}+nm_{2}N_{\theta } \right )$, where $n$ is the total number of test cases, $m_{1}$ is the number of sample mutations, $m_{2}$ is the number of model mutations and $N_{\theta}$ is the number of parameters in the model mutation.
Since $m_{1}$, $m_{2}$ and $N_{\theta }$ are usually very large, PRIMA has to take higher time complexity than activation-based methods ($O(n\log_{}{n})$).

Based on the above analysis, the existing test case prioritization methods show limitations in two aspects:
\textit{(\romannumeral1)} \textit{Limited Scenarios.} Model activation-based prioritization methods have limited application scenarios. Specifically, confidence-based prioritization methods consider that misclassified cases have multiple similar probabilities, thus they are less effective for adversarial cases with high confidence. On the other hand, embedding-based prioritization methods are less effective for FP cases, due to that the embedding features of FP cases are confused with the wrong classes.
\textit{(\romannumeral2)} \textit{High Complexity.} Mutation-based prioritization has a high time complexity since most of them require numerous mutations, memory read and write, and mutation query operations. Therefore, its time complexity is higher than activation-based methods.

\begin{figure}[htbp]
  \centering
  \subfigure[Normal case ``9'']
  {
    \includegraphics[height=0.33\textwidth]{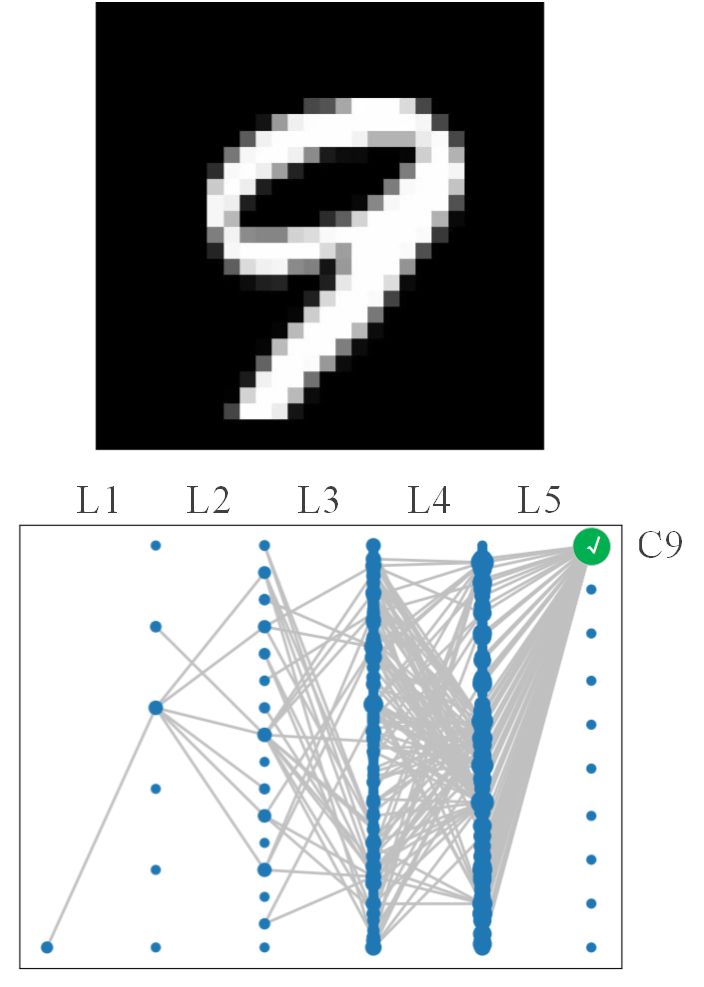}
  }
  \subfigure[Normal case ``7'']
  {
    \includegraphics[height=0.33\textwidth]{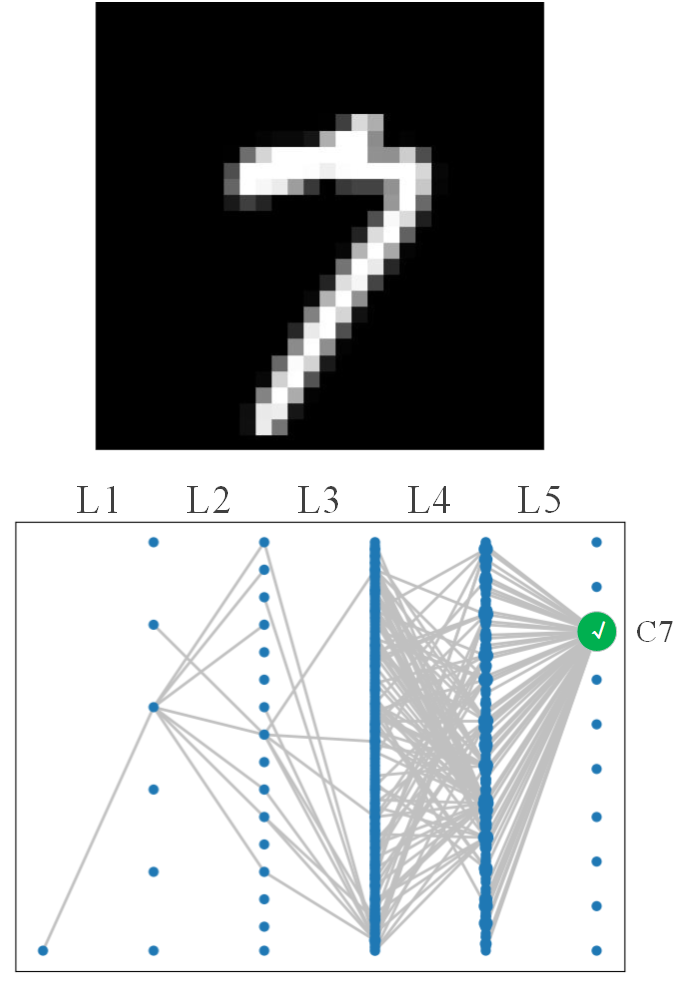}
  }
  \subfigure[FP case ``9'']
  {
    \includegraphics[height=0.33\textwidth]{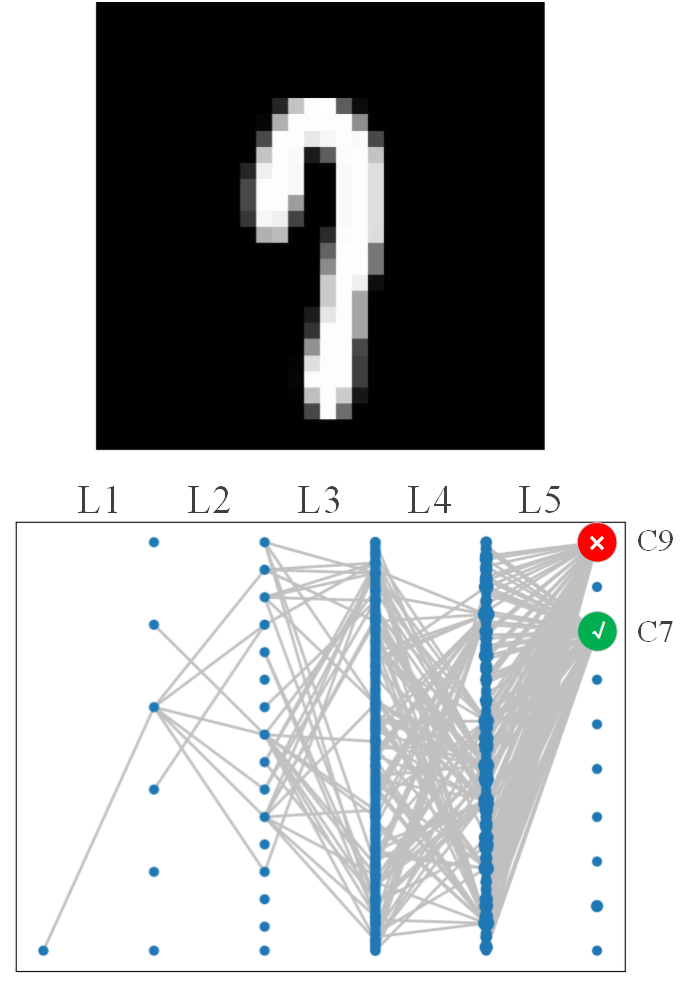}
  }
  \subfigure[Adversarial case ``9'']
  {
    \includegraphics[height=0.33\textwidth]{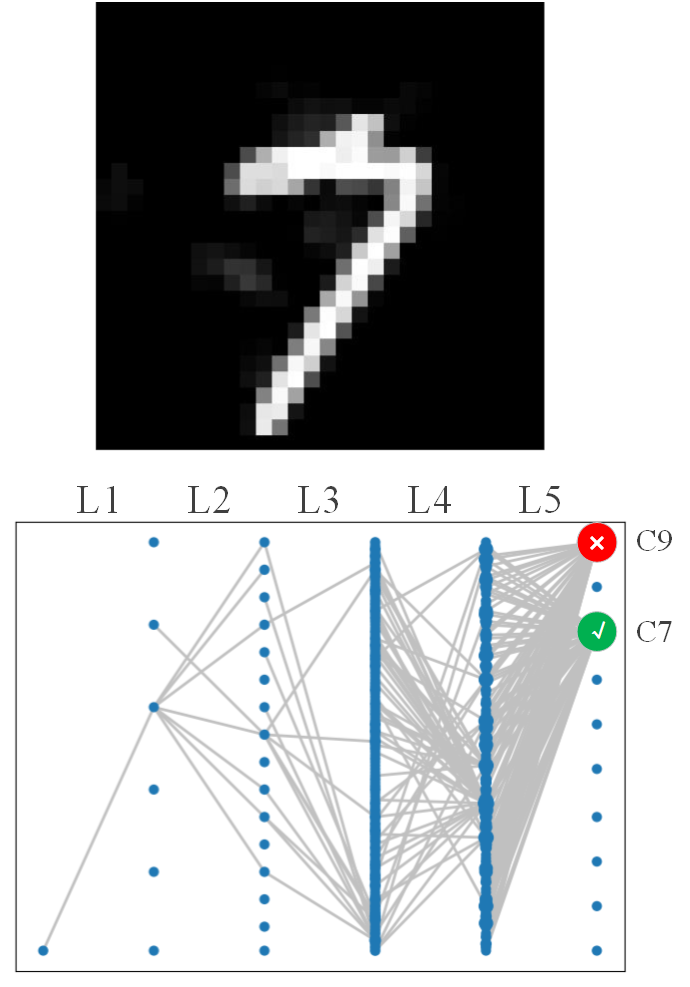}
  }
  \caption{The activation graphs of three type cases. (a) Activation graph of the normal digital 9. (b) Activation graph of the normal digital 7. (c) Activation graph of the false positive digital 7, but misclassified as 9. (d) Activation graph of the  C\&W digital 7, but misclassified as 9.
  }
  \label{fig:ActPath}
\end{figure}

Therefore, to address the prioritization challenges, we explore the in-depth relationship between the test cases and the model's dynamic features. Although the existing model activation-based prioritization methods have extracted features in the hidden layer or the confidence layer, but they overlook some more fine-grained features, such as the neuron's activation.
Therefore, we propose graph-level neuron activation features for test cases, to extract the activation graph between DNN layers.
The activation graph is defined as the connection relationship of neurons.
We study the differences in the activation graph for test cases. 
An illustration is carried out shown in Fig.~\ref{fig:ActPath}. The activation graphs for the normal, FP, and C\&W adversarial test cases are shown.
It is a LeNet-5~\cite{1998MNISTlenet} trained on MNIST~\cite{1998MNISTlenet}.
Fig.~\ref{fig:ActPath} (a) and (b) are normal test cases with class labels 9 and 7, respectively. Fig.~\ref{fig:ActPath} (c) and (d) are misclassified as 9, which are FP and adversarial cases, respectively.
We only show weighted edge values greater than 0.4, in order to show the activation graph clearly. The size of the node is determined by its node feature value.
We observe the distribution differences in activation graphs for the test cases.
Specifically, the edges of the last layer of the activation graph of the normal cases are only connected to the correct node, as shown in Fig.~\ref{fig:ActPath} (a) and (b).
On the contrary, the misclassified cases will connect not only the correct node but also the wrong node in the last layer of the activation graph, as shown in Fig.~\ref{fig:ActPath} (c) and (d).
Comparing the activation graph of normal case (Fig.~\ref{fig:ActPath} (b)) and adversarial case (Fig.~\ref{fig:ActPath} (d)), we observe that the similar distribution of edges in shallow layers (L1, L2), but the differences distribution of edges gradually increase in deeper layers (L3, L4 and L5).

Therefore, we propose a model activation graph-based test case prioritization, namely \textbf{ActGraph}. ActGraph regards the neurons of DNN as nodes, and the adjacency matrix as the connection relationship between nodes. Then, the node features and the adjacency matrix are aggregated by message passing~\cite{2017message}, the aggregated node features will contain the features of neighbor nodes and the structural information between nodes, which can be effectively used for test case prioritization. ActGraph has the following two key characteristics.

\textit{(\romannumeral1)} \textit{Effectiveness and generalizability.} ActGraph extracts the finer-grained activation features of the test cases on the model, and converts the model activations into the spatial relationship of neurons, which can solve the limitation problem of the scene based on the model activation method. ActGraph can prioritize multiple types of test cases by learning one type of cases, which is more general than model activation-based methods.

\textit{(\romannumeral2)} \textit{Efficiency.} ActGraph uses the graph to extract high-level node features instead of mutation operations, which is much more efficient than the mutation-based approaches.

In addition, ActGraph builds a ranking model using Learning-to-Rank (L2R)~\cite{2010Learning}, which can effectively prioritize test cases by learning the node features of test cases. According to the priority results of the ActGraph, the test cases that trigger the vulnerability can be marked earlier, thereby greatly improving the efficiency of DNN testing and effectively saving development time.

The main contributions are summarized as follows.

\begin{itemize}
    \item By identifying the limitations of existing prioritization methods, we propose the activation graph, and find that cases that trigger model vulnerabilities and normal cases in the activation graph.

    \item Motivated by the observation, we propose a novel test case prioritization method, namely ActGraph. It extracts the spatial relationship between neuron nodes, calculates the center node feature in the activation graph, and adopts L2R model to intelligently combine center node feature to achieve efficient test case prioritization.

    \item Comprehensive experiments have been conducted on three datasets and four models to verify the effectiveness and efficiency of ActGraph. It outperforms the SOTA baselines in both natural and adversarial scenarios, especially in \textit{RAUC-100} ($\sim\times$1.40).
\end{itemize}

The remainder of this paper is organized as follows. We describe the related work in Section~\ref{ralated work}. In Section~\ref{approach}, we describe the designed ActGraph method in detail. Experimental results are provided in Section~\ref{experiment}. Threats to validity are described in Section~\ref{threat} and conclusions are made in Section~\ref{conclusions}.

\section{RELATED WORKS\label{ralated work}}

\subsection{Test Case Prioritization for DNNs}
Some DNN test case prioritization methods are proposed to solve the labeling-cost problem. According to the different dynamic features of DNNs for test cases, they can be categorized into: NC-based~\cite{2017DeepXplore, 2018DLFuzz, 2018TensorFuzz, 2017DeepTest, 2018DeepGauge}, model activation-based~\cite{2020DeepGini, 2020Multiple, 2019Guiding, 2019Input} and mutation-based~\cite{2021Prioritizing} test case prioritization methods.

\textit{NC-based test case prioritization.} NC borrows the concept of test coverage from traditional software testing to measure the adequacy of test cases, which was first proposed by Pei \textit{et al}.~\cite{2017DeepXplore}. They argue that DNNs with higher coverage are more secure and reliable when the input is uncertain. Based on this, Guo \textit{et al}.~\cite{2018DLFuzz} continuously mutate the input slightly, with maximizing the NC and the predicted difference between the original and the mutated input to guide the generation of test cases. But some studies~\cite{2019Structural, 2020Is, 2019There} have pointed out that NC and misclassified inputs do not have a strong correlation, so cannot be effectively applied to test case prioritization.

\textit{Model activation-based test case prioritization.} Test case prioritization methods based on model activation can be divided into confidence-based and embedding feature-based. Confidence-based prioritization methods generally obtain the confidence output of the DNN. From a statistical point of view, they assume that correctly classified test cases should have high probability, and misclassified or abnormal test cases have multiple similar probabilities. For example, Feng \textit{et al}.~\cite{2020DeepGini} proposed DeepGini based on the Gini impurity, which can quickly identify test samples that may lead to DNN misclassification. Shen \textit{et al}.~\cite{2020Multiple} proposed Multiple-Boundary Clustering and Prioritization (MCP) to uniformly select high-priority test cases from all boundary regions by dividing test cases into multiple boundary regions. Embedding feature-based prioritization methods believe that the deep layer of the DNN can better represent the high-level features of the input cases, so the last layer of hidden layer features of the DNN output is generally obtained. Kim \textit{et al}.~\cite{2019Guiding} proposed Surprise Adequacy (SA) to measure the adequacy of test cases for DNN testing by calculating the difference in hidden layer features between test cases and training data, including Likelihood-based Surprise Adequacy (LSA) and Distance-based Surprise Adequacy (DSA). LSA refers to estimating the density of the embedding features of the test cases on the training data, while DSA defines the Euclidean distance between the embedding features of the test cases and the training data. They show that DSA is more suitable for classification models.

\textit{Mutation-based test case prioritization.} Wang \textit{et al}.~\cite{2021Prioritizing} proposed PRIMA, which prioritizes those test cases that produce different prediction results through fuzzy mutation operations (including input mutation and model mutation), arguing that this is more likely to reveal DNN vulnerabilities.

In addition, some test case sampling methods are proposed to improve the efficiency of DNN testing~\cite{2019Boosting, 2020Practical, 2021Test, 2019Behavior, 2020Operational}, to estimate the accuracy of DNN by sampling a few test cases. Our work aims to identify more bug-revealing test cases earlier by prioritizing test cases.

\subsection{Graph Structure of DNNs}
Several studies have constructed DNNs as Graphs (DAG) to explore the performance of DNNs, such as interpretability, generalization and performance analysis.
Naitzat \textit{et al}.~\cite{2020Topology} demonstrated the superiority of ReLu activation by studying the variation of the Betty number of two classes of DNN.
Filan~\cite{filan2021clusterability} \textit{et al}. proposed to directly represent the fully connected layer as a weighted undirected graph, where each neuron corresponds to a node.
Rieck~\cite{2019Neural} \textit{et al}. proposed neural persistence, a measure of the topological complexity of network structures that can give a criterion for early stopping of training. They proposed "Unroll", and converting convolutional layers into graphs.
Then, Vahedian \textit{et al}.~\cite{2021Convolutional} proposed a “Rolled” graph representation of convolutional layers to solve the DNN performance prediction problem by capturing the early DNN dynamics during the training phase. To maintain the semantic meaning of the convolutional layers, they represent each filter as a node and link the filters in successive layers by weighted edges.
Zhao \textit{et al}.~\cite{2021Quantitative} proposed feature entropy to quantitatively identify individual neuron states of a specific class.

In general, most of the existing DAG studies focus on the relationship between the structural parameters of the DNN and the overall behavior of the DNN. ActGraph and the existing DAG mainly have the following differences: \textit{\romannumeral1)} \textit{The methods of constructing the graph are different.} The graph of ActGraph uses the DNN structure and multi-layer activation, unlike past works that only consider the DNN structure; \textit{\romannumeral2)} \textit{The application scenarios are different.} ActGraph is used for test case prioritization in DNN testing.









\section{APPROACH\label{approach}}
\subsection{Overview}
Existing DAG studies have been able to convert the DNN into the graph, discussing the dynamic properties of the DNN model~\cite{filan2021clusterability, 2019Neural, 2021Convolutional, 2021Quantitative}. They used an undirected weighted graph to treat neurons as nodes and the weights of the model as weighted edges between nodes. But they did not consider expressing the activation information of case to graph. Different from them, we use a directed weighted graph to take the model weights as the skeleton of the graph, and map the activation values to the graph, so that the feature of the test case is expressed on the graph.
For convenience, the definitions of symbols used in this paper are listed in Table~\ref{symbol}.

\begin{table}[htb]
  \centering
  \caption{The definitions of symbols.}
  \resizebox{0.6\linewidth}{!}{
    \begin{tabular}{c|r}
    \hline
    \textbf{Symbol}                                                                                            & \textbf{Definition}                                                   \\ \hline
    $x$                                                                                                        & a test case                                                           \\
    $n_{i}^{l}$                                                                                                & the $i$-th neuron of the $l$-th layer                                 \\
    $F_{i}^{l}$                                                                                                & the feature map of $n_{i}^{l}$                                        \\
    $\varphi_{i}^{l}$                                                                                          & the average and normalized value of $F_{i}^{l}$                       \\
    $\theta$                                                                                                   & the model weights of the trained DNN                                  \\
    $W$                                                                                                        & the average and normalized weights of $\theta$                        \\
    $D=(V,E)$                                                                                                  & the direct weighted graph with sets of nodes and edges                \\
    $v_{i}$                                                                                                    & the $i$-th node of $D$                                                \\
    $\left \langle v_{j},v_{i} \right \rangle $                                                                & the directed edge from $v_{j}$ to $v_{i}$                             \\
    $\Gamma_{D}^{-}(v_{i})$                                                                                    & the set of predecessors of $v_{i}$                                    \\
    $A$                                                                                                        & the adjacency matrix of $D$                                           \\
    $nf$                                                                                                       & the node feature                                                      \\
    $cnf$                                                                                                      & the center node feature                                               \\
    $AGG(\cdot)$                                                                                               & the aggregation function of the message passing                       \\
    $y$                                                                                                        & the flag of whether the $x$ triggers the model vulnerability (0 or 1) \\
    $\Omega(\cdot)$                                                                                            & the regularization function                                           \\
    $T$                                                                                                        & the number of trees of xgboost                                        \\ \hline
    \end{tabular}
  }
  \label{symbol}%
\end{table}%

\begin{figure}[htb]
    \centering
    \includegraphics[width=0.9\textwidth]{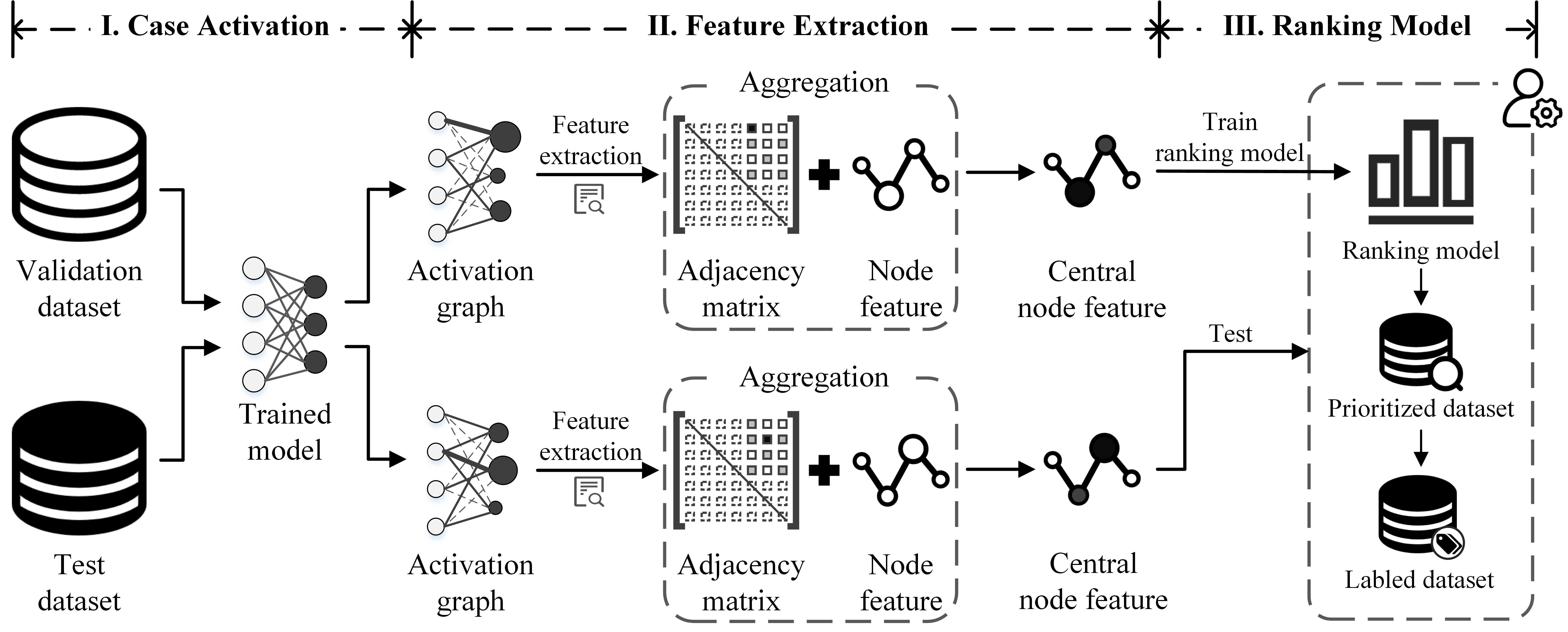}
    \caption{The framework of ActGraph, which be divided into three stages: test case activation, the feature of activation graph extraction and ranking model building.}
    \label{pipline}
\end{figure}

We propose a novel test case prioritization method based on model activation graph, namely ActGraph. It consists of three stages.
\textit{\romannumeral1) Test Case Activation}:
the test cases are fed into the trained DNN, and each layer of the DNN outputs activation values (Section~\ref{test_cases_act});
\textit{\romannumeral2) Feature Extraction}:
activation graphs are constructed based on the activations of DNN's each layer, and the adjacency matrix and node features are extracted on the activation graphs. Finally the center node features are obtained by message passing aggregation (Section~\ref{fea_act_graph_extract});
\textit{\romannumeral3) Ranking Model Building}:
ActGraph adopts the framework of L2R to build a ranking model, which can utilize the center node features, for prioritizing test cases (Section~\ref{rank_model_build}).
The framework of ActGraph is shown in Fig.~\ref{pipline}.

\subsection{Test Case Activation\label{test_cases_act}}
ActGraph is a model activation-based test case prioritization method that runs during the test phase. Test cases are input to the DNN, and each layer outputs activation values.
In order to facilitate the construction of the graph, the weights and activations of each neuron are averaged, and the weights and activations of each layer are normalized.

For a trained DNN and a test case $x$. DNN has $L$ layers, $n_{i}^{l}$ is the $i$-th neuron of the $l$-th layer.
The case $x$ is input to the DNN, and get the output of each layer of neurons in the DNN. The activation value $\varphi _{i}^{l}$ of $n_{i}^{l}$ is calculated as:
\begin{equation}\label{actvalue}
    \centering
    \varphi _{i}^{l} (x)=\frac{1}{Height_{l}\times Width_{l}} \sum_{Height_{l}}^{} \sum_{Width_{l}}^{}F_{i}^{l} (x)
\end{equation}
where $F_{i}^{l}(x) \in \mathbb{R} ^{Height_{l} \times Width_{l} \times Channel_{l}}$ is the feature map of $n_{i}^{l}$ output when $x$ is input. For the convolutional layer, the output dimension is $ Height_{l} \times Width_{l}$, and the dimension of the fully connected layer is $1 \times 1$.

The neuron activation value $\varphi ^{l} (x)$ of each layer performs the max-min normalization, which normalizes $\varphi ^{l} (x)$ to the [0,1] range, is calculated as:
\begin{equation}\label{actvalue_norm}
    \centering
    \varphi ^{l} (x) = \frac{\varphi ^{l} (x)-min(\varphi ^{l} (x))}{max(\varphi ^{l} (x)) - min(\varphi ^{l} (x))}
\end{equation}

For the neuron $n_{i}^{l} $, its neuron weight is calculated as:
\begin{equation}\label{neuron_weight}
    \centering
    w_{j,i}^{l-1,l} =\frac{1}{Height_{l}^{\theta } \times Width_{l}^{\theta }} \sum_{Height_{l}^{\theta }}^{} \sum_{Width_{l}^{\theta }}^{}\theta _{j,i}^{l-1,l}
\end{equation}
where $\theta _{j,i}^{l-1,l}  \in \mathbb{R} ^{Height_{l}^{\theta } \times Width_{l}^{\theta }}$ represents the weight parameter between the neuron $n_{j}^{l-1} $ and the neuron $n_{i}^{l} $. The dimension of the neuron weight of the convolutional layer of the $l$-th layer is $Height_{l}^{\theta } \times Width_{l}^{\theta } $, and the dimension of the weight of the fully connected layer is $1 \times 1$.

Normalize the neuron weight $w_{}^{l-1,l} $ of each layer, and convert $w_{}^{l-1,l} $ of each layer to the range of [0,1], is calculated as:
\begin{equation}\label{neuron_weight_norm}
    \centering
    w_{}^{l-1,l} = \frac{w_{}^{l-1,l} - min(w_{}^{l-1,l} )}{max(w_{}^{l-1,l} ) - min(w_{}^{l-1,l} )}
\end{equation}

To reduce the computational cost, ActGraph only obtains the neuron activations and their weights of the last $K$ layers of the DNN. 

\subsection{Feature Extraction\label{fea_act_graph_extract}}
ActGraph adopts the L2R framework, and builds a ranking model for each DNN model. Since L2R requires a set of features like other supervised machine learning~\cite{2010Learning}. ActGraph extracts a set of features from the activation values of the test cases and the structural information of the model. In this section, we propose the steps of feature extraction for test cases in ActGraph.

As shown in Fig.~\ref{fig:ActPath}, the weighted edges can significantly represent the differences of distribution between different test cases, but cannot clearly express the characteristics of neurons.
Therefore, we would like to extract more effective node features from the activation graph for prioritization. The message passing of Graph Neural Network (GNN) can aggregate the features of the current node and neighbor nodes, which is similar to the data flow of DNN, that is, the activation values of the previous layer are passed to the next layer. Specifically, we use a directed activation graph, and extract the weighted in-degrees of nodes as node features. The weighted in-degrees are low-order node features that represent the importance of the nodes. Further, we aggregate node features and the adjacency matrix to obtain higher-order node features by message passing, namely center node feature. For the explanation of its effectiveness, we describe it in detail in Section~\ref{rq3}.


Because the activation values of the DNN have the data flow, we construct the DNN as a directed weighted graph. Let $D=(V,E)$ be a directed weighted graph whose node set is $V$ and its edge set is $E$, where $V$ is the neuron set of DNN, and $E$ is the set of the directed weighted edges, as follows:
\begin{equation}\label{activation_graph}
    A_{j,i}=\left\{\begin{matrix}
     w_{j,i} \times \varphi_{i} & , v_{j}\in \Gamma_{D}^{-}(v_{i} )   \\
     0 & ,v_{j}\notin \Gamma_{D}^{-}(v_{i} )
    \end{matrix}\right.
\end{equation}
where $A$ is the adjacency matrix of $D$, $v_{i}$ is the $i$-th node of $D$, $w_{j,i}$ is the weight between $v_{j}$ and $v_{i}$, and $\Gamma_{D}^{-}(v_{i}) = \left \{ v_{j} \mid v_{j}\in V(D) \wedge \left \langle v_{j},v_{i} \right \rangle \in E(D) \wedge v_{j}\ne v_{i} \right \}$ is the set of predecessors of $v_{i}$.

We use weighted in-degree as node features ($nf$). Degree is the simplest and most effective feature for nodes, which captures the connectivity of nodes. The $v_{i}$'s weight is the sum of the weights of adjacent input edges, calculated as follows:
\begin{equation}\label{node_feature}
    nf_{i} =\begin{matrix}
    \sum_{j}^{}A_{j,i}    & ,v_{j} \in \Gamma_{D}^{-}(v_{i})
    \end{matrix}
\end{equation}
where $nf_{i}$ is the node feature value of $v_{i}$. Weighted in-degree is a low-order node feature. Therefore, we use the message passing of GNN to aggregate the adjacency matrix and node features of the activation graph to obtain the center node feature (\textit{cnf}), which is calculated as follows:
\begin{equation}\label{central_node_feature}
    cnf= \begin{matrix}
    AGG(A_{}^{},nf)  
    \end{matrix}
\end{equation}
where the aggregation function $AGG()$ can use $Sum()$, $Max()$, and $Average()$. We use the $Sum()$ function.

After calculating the \textit{cnf} of all nodes, ActGraph only takes the \textit{cnf} of the last two layers. Because we believe that the deeper activation of the model can fully express the high-dimensional characteristics of test cases. The two-layer \textit{cnf} needs at least four layers of weight and activation, so we set $K$=4. The reason is described in Section~\ref{proof}.

\subsection{Ranking Model Building\label{rank_model_build}}
ActGraph adopts the XGBoost algorithm~\cite{2016XGBoost}, which is an optimized distributed gradient reinforcement learning algorithm, and establishes an L2R-based ranking model.

The \textit{cnf} of the validation set obtained by Eq.~(\ref{central_node_feature}) is used as the training set of the ranking model, and according to the DNN's prediction of the sample, it is labeled as 0 (prediction is correct) or 1 (prediction is wrong). The loss function for training the ranking model is as follows:
\begin{equation}\label{ranking model}
   obj(cnf, y)=l(y, \hat{y})+\sum_{t=1}^{T}\Omega  (f_{t})
\end{equation}
where $\hat{y}=\sum_{t=1}^{T} f_{t}(cnf)$, $f_{t}(cnf)$ is the predicted value of the $t$-th tree, $y$ is 0 or 1, $T$ is the number of trees, and $\Omega $ is regularization.
In summary, the process of training ranking model is shown in Algorithm~\ref{algorithm1}.

\begin{algorithm}[t]
\caption{ActGraph}
\begin{algorithmic}[1]

\renewcommand{\algorithmicrequire}{\textbf{Input:}}
\REQUIRE 
A DNN $f_{1}$ to be tested; The last $K$ layers selected; Validation dataset $x_{c} \in X=\{x_{1}, x_{2}, ...\}$; The set of center node feature $cnf$.

\renewcommand{\algorithmicensure}{\textbf{Output:}}
\ENSURE A ranking model $f_{2}$. 

\FOR{$l$ in $K$}
    \STATE Obtain neuron weight $w^{l-1,l}$ by Eq.~(\ref{neuron_weight}).
    \STATE Normalize the neuron weight $w^{l-1,l}$ by Eq.~(\ref{neuron_weight_norm}).
\ENDFOR
\STATE $cnf = \left \{ \emptyset \right \}$.
\FOR{$x_{c}$ in $X$}
    \FOR{$l$ in $K$}
        \STATE Obtain neuron activation $\varphi ^{l} (x_{c} )$ by Eq.~(\ref{actvalue}).
        \STATE Normalize the neuron activation $\varphi ^{l} (x_{c} )$ by Eq.~(\ref{actvalue_norm}).
    \ENDFOR
    \STATE Initialize directed weighted graph $D_{c}$.
    \STATE Extract Adjacency matrix $A_{c}$ by Eq.~(\ref{activation_graph}).
    \STATE Calculate node feature $nf_{c}$ by Eq.~(\ref{node_feature}).
    \STATE Aggregate center node feature $cnf_{c}$ by Eq.~(\ref{central_node_feature}).
    \STATE $cnf \gets cnf \cup cnf_{c}$.
\ENDFOR
\STATE Train the ranking model $f_{2}$ by Eq.~(\ref{ranking model}).
\RETURN $f_{2}$.
\end{algorithmic}
\label{algorithm1}
\end{algorithm}

\subsection{Utility Analysis of Center Node Feature\label{proof}}
In this section,
we analyze the utility of \textit{cnf} and explain how the $K$ value of ActGraph is determined. Let a DNN with $N$ neurons. Its output value of each layer is activated by a case $x$. The activation value $\varphi$ and the weight $W$ can be expressed as:
\begin{equation}
\label{activationspecial}
    \varphi =\begin{bmatrix}
    \varphi_{0}(x) & \varphi_{1}(x) & ... & \varphi_{N-1}(x)
    \end{bmatrix}_{1\times N}
\end{equation}
\begin{equation}
\label{weight_special}
    W=\begin{bmatrix}
    w_{0,0} & w_{0,1} & ... & w_{0,N-1}\\
    w_{1,0} & w_{1,1} & ... & w_{1,N-1}\\
    ... & ... & ... & ...\\
    w_{N-1,0} & w_{N-1,1} & ... & w_{N-1,N-1}
    \end{bmatrix}_{N\times N}
\end{equation}
where $\varphi$ and $W$ are layer normalized by Eq.~(\ref{actvalue_norm}) and Eq.~(\ref{neuron_weight_norm}). Then the adjacency matrix $A$ is calculated by Eq.~(\ref{activation_graph}) as:
\begin{equation}
\label{AP_special}
    \small
    A = \begin{bmatrix}
    \varphi_{0}\cdot w_{0,0} & \varphi_{1}\cdot w_{0,1} & ... & \varphi_{N-1}\cdot w_{0,N-1}\\
    \varphi_{0}\cdot w_{1,0} & \varphi_{1}\cdot w_{1,1} & ... & \varphi_{N-1}\cdot w_{1,N-1}\\
    ... & ... & ... & ...\\
    \varphi_{0}\cdot w_{N-1,0} & \varphi_{1}\cdot w_{N-1,1} & ... & \varphi_{N-1}\cdot w_{N-1,N-1}
    \end{bmatrix}_{N\times N}
\end{equation}

Then the node feature $nf$ is calculated by Eq.~(\ref{node_feature}) as:
\begin{equation}
\label{NF_special}
    nf=\begin{bmatrix}
    \varphi_{0}\sum_{}^{j}w_{j,0}  & ... & \varphi_{N-1}\sum_{}^{j}w_{j,N-1}
\end{bmatrix}_{1\times N}
\end{equation}
where $nf$ is the in-degree of the activation graph. A node feature $nf_{i}$ is $\varphi_{i}\sum_{}^{j}w_{j,i}$, which indicates that the $nf_{i}$ value is obtained by aggregating the activation value and input edges of $v_{i}$. Finally, the center node feature \textit{cnf} is calculated by computing $A$ and $nf$ by Eq.~(\ref{central_node_feature}) as:
\begin{equation}
\label{CNF_special}
    cnf=\begin{bmatrix}
     \sum_{}^{z}\varphi_{z}w_{z,0}nf_{z} & ... & \sum_{}^{z}\varphi_{z}w_{z,N-1}nf_{z}
    \end{bmatrix}_{1\times N}
\end{equation}
where the $cnf_{i}$ of $v_{i}$ is $\sum_{}^{z}\varphi_{z}w_{z,i}nf_{z}$. Intuitively, $cnf_{i}$ is aggregated by the activation values and the $nf$ value of neurons in the upper layer of $v_{i}$.

Therefore, ActGraph requires at least three layers of network to aggregate so that the \textit{cnf} of the last layer is valid (not zero). In the experiment, we use the last four layers of DNN, i.e. $K$=4, to obtain the effective \textit{cnf} of the last two layers.

\section{EXPERIMENT\label{experiment}}
We evaluate ActGraph through answering the following five research questions (RQs).
\begin{itemize}
    \item \textbf{RQ1}: Does ActGraph show the SOTA prioritization performance in both natural and adversarial scenarios?
    \item \textbf{RQ2}: Does ActGraph show the competitive generalizability in mixed scenarios?
    \item \textbf{RQ3}: How to interpret ActGraph's utility by t-SNE and heatmap visualization?
    \item \textbf{RQ4}: How is the stability of ActGraph under different hyperparameters (i.e. trainset size and training parameters of ActGraph)?
    \item \textbf{RQ5}: Does ActGraph efficient in time complexity?
\end{itemize}

\subsection{Setup}\label{setup}
\textbf{Platform.}
The experiments were conducted on a server equipped with Intel XEON 6240 2.6GHz X 18C (CPU), Tesla V100 32GiB (GPU), 16GiB DDR4-RECC 2666MHz (Memory), Ubuntu 16.04 (OS), Python 3.6, Keras-2.2.4, Tensorflow-gpu-1.9.0, Xgboost-1.5.2.

\textbf{Datasets.}
We conduct experiment on MNIST~\cite{1998MNISTlenet}, CIFAR-10~\cite{2009cifar} and CIFAR-100~\cite{2009cifar}. MNIST contains 60,000 28$\times$28 gray-scale images, and each image is marked with numbers from 0 to 9. CIFAR-10 includes 60,000 32$\times$32 three-channel RGB-color images, which are divided into ten classes equally. CIFAR-100 includes 60,000 32$\times$32 three-channel RGB-color images, which are divided into one hundred classes equally. For each dataset, 40,000 are used for training, 10,000 for validation and 10,000 for testing.

\textbf{Models.}
For MNIST, we use LeNet-5~\cite{1998MNISTlenet} for prioritization. On CIFAR-10, VGG16~\cite{2014vgg} and ResNet18~\cite{2015resnet} are adopted. On even larger dataset CIFAR-100, we adopt VGG19~\cite{2014vgg} model. The datasets and models configurations are shown in Table~\ref{config1}.

\begin{table}[htbp]
  \centering
  \caption{Dataset and model configurations.}
  \resizebox{0.7\linewidth}{!}{
\begin{tabular}{ccccccc}
\hline
\textbf{Datasets}         & \textbf{Categories} & \textbf{Training data}  & \textbf{Validation data} & \textbf{Models} & \textbf{Params} & \textbf{Testing acc} \\ \hline
MNIST                     & 10                  & 40,000                  & 10,000                   & LeNet-5         & 107,786         & 99.20\%              \\ \hline
\multirow{2}{*}{CIFAR-10} & \multirow{2}{*}{10} & \multirow{2}{*}{40,000} & \multirow{2}{*}{10,000}  & VGG16           & 134,326,366     & 92.04\%              \\
                          &                     &                         &                          & ResNet18        & 273,066         & 92.56\%              \\ \hline
CIFAR-100                 & 100                 & 40,000                  & 10,000                   & VGG19           & 139,638,622     & 70.08\%              \\ \hline
\end{tabular}
  }
  \label{config1}%
\end{table}%

\textbf{Data Preparation.}
To verify that ActGraph is able to prioritize various test cases that trigger model bugs. We use a variety of data operations for data generation, and generate a variety of different types of datasets to make the DNN misclassify, including adversarial and natural noise cases. The natural operations include image rotation, translation and flipping, collectively referred to as \textbf{Rotate}. And the adversarial operations include C\&W~\cite{2017CW} and Jacobian-based Saliency Map Attacks (JSMA)~\cite{2015JSMA}. C\&W attack leads the model to output false label with high confidence. JSMA can change only a few pixels to implement the attack, so the power of the disturbance is small.

We construct the \textit{Testset} to prioritize and the \textit{Trainset} for training the ranking model. For \textbf{Original}, \textit{Testset} comes from the testset of the original dataset, and \textit{Trainset} comes from the validation set of the original dataset. However, the DNN has high accuracy and only a few misclassified samples in \textit{Trainset}. Therefore, these samples need to be repeatedly sampled until the balance of positive and negative samples reaches 5,000 to 5,000, and \textit{Testset} remains unchanged. For other types of data (\textbf{Rotate}, \textbf{JSMA}, \textbf{C\&W} and \textbf{Mix}), the ratio of \textit{Trainset} is 5,000 to 5,000, that is, it consists of 5,000 normally classified samples and 5,000 manipulated misclassified samples, and the ratio of \textit{Testset} is 8,000 to 2,000. Mix is randomly sampled from four types of sets.

\textbf{Baselines.}
We adopt the model activation-based and mutation-based prioritization algorithms as the baselines in our experiment, including DeepGini~\cite{2020DeepGini}, MCP~\cite{2020Multiple}, DSA~\cite{2019Guiding} and PRIMA~\cite{2021Prioritizing}. The parameters for these algorithms are configured following their settings reported in the respective papers. In addition, in order to explore the impact of ActGraph extracted graph-level features on test case prioritization, we extracted and concatenated the confidence output and the last hidden output as our baseline, namely Act.

\textbf{Metrics.}
We use RAUC~\cite{2021Prioritizing} as the evaluation metric for prioritization. RAUC is calculated as the ratio of the area under the curve for the test input prioritization approach to the area under the curve of the ideal prioritization, as follows:

\begin{equation}\label{rauc}
    \centering
    RAUC=\frac{AUC(rank)}{AUC(ideal\_rank)}\times 100\%
\end{equation}

\textit{RAUC-n} is the RAUC for the first $n$ test cases, as follows:

\begin{equation}\label{rauc-n}
    \centering
    RAUC(n)=\frac{AUC(rank,n)}{AUC(ideal\_rank,n)}\times 100\%
\end{equation}

The value range is [0, 1], and 1 indicates the best ranking result. \textit{RAUC-100}, \textit{RAUC-500}, \textit{RAUC-1000} and \textit{RAUC-ALL} are used in the experiments.

\textbf{Implementation Details.}
Our experiments have the following settings:
(1) For XGBoost ranking algorithm in ActGraph, we set \textit{$Learning\_rate$} to be 0.1, \textit{$Colsample\_bytree$} to be 0.3, and \textit{$Max\_depth$} to be 5;
(2) In order to reduce the computational cost, we set $K$ as 4 and take the \textit{cnf} of the last two layers;
(3) For all image data, we normalize the range of each pixel to [0, 1].

\subsection{Effectiveness of ActGraph}\label{rq1}
In the section, we find the answer to \textbf{RQ1}, by comparing ActGraph with 5 baseline algorithms to verify the effectiveness of ActGraph in natural and adversarial scenarios.

\textbf{Implementation details.}
(1) Each model and dataset is set with two natural scenarios (Original and Rotate) and two adversarial scenarios (C\&W and JSMA).
(2) The training type is the same as the test type.
(3) The size of trainset of ranking model is 2,000, which contains 1,000 positive samples and 1,000 negative samples.

\textbf{Results and analysis.}
The results are shown in Table~\ref{rq1table}. The bold indicates the optimal results of different methods under the same type scenario and the same metric.
In the total 64 results, ActGraph performs the best with 42 best results (65.63\%), followed by DeepGini with 13 best results (20.31\%) and PRIMA with 7 best results (10.94\%).
Then, we average the four metrics, in which ActGraph is the best, followed by PRIMA. Specifically, the average results of ActGraph are 0.865$\sim$0.939, which are 0.80\%$\sim$5.96\% higher than PRIMA, 2.06\%$\sim$13.19\% higher than DeepGini, 8.53\%$\sim$13.50\% higher than Act, 18.78\%$\sim$21.28\% higher than MCP, and 11.14\%$\sim$24.58\% higher than DSA.
In the 32 results of natural scenarios, ActGraph gets 17 best results, DeepGini gets 13 best results, and PRIMA gets 2 best results.
In the 32 results of adversarial scenarios, ActGraph gets 25 best results, PRIMA gets 5 best results, MCP gets 1 best results and DSA gets 1 best results.

Because the time and cost of prioritization is limited, the number of test cases that can be labeled is often small. This also means that \textit{RAUC-100} is more important than \textit{RAUC-ALL} for the test case prioritization approaches.
In \textit{RAUC-100}, the average result of ActGraph is 0.871, 13.19\% higher than DeepGini, 5.96\% higher than PRIMA, 20.20\% higher than MCP, 24.36\% higher than DSA, and 13.50\% higher than Act.
These results show that DeepGini has better effects than PRIMA in natural scenarios, PRIMA has better effects than DeepGini in adversarial scenarios, and the average results of PRIMA are better than DeepGini. ActGraph shows the SOTA effect in both adversarial and natural scenarios, especially in \textit{RAUC-100}.

\begin{table*}[t]
  \centering
  \caption{The results of prioritization in adversarial and natural scenarios.}
  \resizebox{1\linewidth}{!}{
    \Large
\begin{tabular}{ccccccccccccccccccc}
\hline
\multicolumn{2}{c|}{\textbf{Datasets}}                  & \multicolumn{4}{c|}{MNIST}                                                                         & \multicolumn{8}{c|}{CIFAR-10}                                                                                                                                                                                       & \multicolumn{4}{c|}{CIFAR-100}                                                                           & \multirow{3}{*}{Average} \\ \cline{1-18}
\multicolumn{2}{c|}{\textbf{Models}}                    & \multicolumn{4}{c|}{LeNet-5}                                                                       & \multicolumn{4}{c|}{VGG16}                                                                               & \multicolumn{4}{c|}{ResNet18}                                                                            & \multicolumn{4}{c|}{VGG19}                                                                               &                          \\ \cline{1-18}
\multicolumn{2}{c|}{\textbf{Type}}                      & Original       & Rotate               & JSMA                 & \multicolumn{1}{c|}{CW}             & Original             & Rotate               & JSMA                 & \multicolumn{1}{c|}{CW}             & Original             & Rotate               & JSMA                 & \multicolumn{1}{c|}{CW}             & Original             & Rotate               & JSMA                 & \multicolumn{1}{c|}{CW}             &                          \\ \hline
\multirow{6}{*}{R-100}  & \multicolumn{1}{c|}{DeepGini} & 0.535          & 0.967                & 0.954                & \multicolumn{1}{c|}{0.906}          & \textbf{0.653}       & 0.794                & 0.869                & \multicolumn{1}{c|}{0.639}          & \textbf{0.586}       & 0.845                & 0.888                & \multicolumn{1}{c|}{0.559}          & 0.828                & 0.822                & 0.551                & \multicolumn{1}{c|}{0.430}          & 0.739                    \\
                        & \multicolumn{1}{c|}{PRIMA}    & 0.386          & 0.943                & 1.000                & \multicolumn{1}{c|}{0.984}          & 0.530                & 0.765                & 0.892                & \multicolumn{1}{c|}{0.983}          & 0.415                & 0.739                & 0.889                & \multicolumn{1}{c|}{\textbf{0.989}} & \textbf{0.886}       & 0.864                & 0.766                & \multicolumn{1}{c|}{0.948}          & 0.811                    \\
                        & \multicolumn{1}{c|}{MCP}      & 0.279          & 0.945                & 0.870                & \multicolumn{1}{c|}{0.844}          & 0.369                & 0.793                & 0.931                & \multicolumn{1}{c|}{0.963}          & 0.489                & 0.800                & 0.917                & \multicolumn{1}{c|}{0.923}          & 0.400                & 0.432                & 0.379                & \multicolumn{1}{c|}{0.371}          & 0.669                    \\
                        & \multicolumn{1}{c|}{DSA}      & 0.160          & 0.877                & 0.854                & \multicolumn{1}{c|}{0.926}          & 0.428                & 0.582                & 0.712                & \multicolumn{1}{c|}{0.582}          & 0.262                & 0.495                & 0.684                & \multicolumn{1}{c|}{0.664}          & 0.586                & 0.536                & \textbf{0.834}       & \multicolumn{1}{c|}{0.854}          & 0.627                    \\
                        & \multicolumn{1}{c|}{Act}      & 0.216          & 0.945                & 0.944                & \multicolumn{1}{c|}{0.891}          & 0.369                & 0.830                & 0.893                & \multicolumn{1}{c|}{0.944}          & 0.526                & 0.929                & 0.862                & \multicolumn{1}{c|}{0.671}          & 0.700                & 0.724                & 0.599                & \multicolumn{1}{c|}{0.731}          & 0.736                    \\
                        & \multicolumn{1}{c|}{ActGraph} & \textbf{0.623} & \textbf{0.991}       & \textbf{1.000}       & \multicolumn{1}{c|}{\textbf{1.000}} & 0.519                & \textbf{0.915}       & \textbf{0.957}       & \multicolumn{1}{c|}{\textbf{1.000}} & 0.574                & \textbf{0.956}       & \textbf{0.927}       & \multicolumn{1}{c|}{0.954}          & 0.885                & \textbf{0.933}       & 0.746                & \multicolumn{1}{c|}{\textbf{0.955}} & \textbf{0.871}           \\ \hline
\multirow{6}{*}{R-500}  & \multicolumn{1}{c|}{DeepGini} & 0.672          & \textbf{0.978}       & 0.977                & \multicolumn{1}{c|}{0.946}          & \textbf{0.604}       & 0.827                & 0.879                & \multicolumn{1}{c|}{0.828}          & \textbf{0.550}       & 0.878                & 0.924                & \multicolumn{1}{c|}{0.760}          & 0.852                & 0.858                & 0.636                & \multicolumn{1}{c|}{0.544}          & 0.795                    \\
                        & \multicolumn{1}{c|}{PRIMA}    & 0.634          & 0.943                & 0.994                & \multicolumn{1}{c|}{0.991}          & 0.481                & 0.763                & 0.893                & \multicolumn{1}{c|}{0.976}          & 0.450                & 0.691                & 0.921                & \multicolumn{1}{c|}{\textbf{0.972}} & \textbf{0.855}       & 0.857                & \textbf{0.863}       & \multicolumn{1}{c|}{0.938}          & 0.826                    \\
                        & \multicolumn{1}{c|}{MCP}      & 0.542          & 0.907                & 0.870                & \multicolumn{1}{c|}{0.807}          & 0.451                & 0.774                & 0.927                & \multicolumn{1}{c|}{0.905}          & 0.471                & 0.734                & 0.908                & \multicolumn{1}{c|}{0.862}          & 0.435                & 0.400                & 0.397                & \multicolumn{1}{c|}{0.377}          & 0.673                    \\
                        & \multicolumn{1}{c|}{DSA}      & 0.287          & 0.913                & 0.869                & \multicolumn{1}{c|}{0.880}          & 0.403                & 0.647                & 0.717                & \multicolumn{1}{c|}{0.652}          & 0.222                & 0.475                & 0.615                & \multicolumn{1}{c|}{0.626}          & 0.651                & 0.501                & 0.720                & \multicolumn{1}{c|}{0.722}          & 0.619                    \\
                        & \multicolumn{1}{c|}{Act}      & 0.293          & 0.954                & 0.942                & \multicolumn{1}{c|}{0.912}          & 0.451                & 0.778                & 0.896                & \multicolumn{1}{c|}{0.903}          & 0.456                & 0.934                & 0.826                & \multicolumn{1}{c|}{0.621}          & 0.807                & 0.784                & 0.673                & \multicolumn{1}{c|}{0.662}          & 0.743                    \\
                        & \multicolumn{1}{c|}{ActGraph} & \textbf{0.673} & 0.975                & \textbf{0.999}       & \multicolumn{1}{c|}{\textbf{1.000}} & 0.528                & \textbf{0.852}       & \textbf{0.956}       & \multicolumn{1}{c|}{\textbf{0.990}} & 0.477                & \textbf{0.976}       & \textbf{0.926}       & \multicolumn{1}{c|}{0.970}          & 0.847                & \textbf{0.865}       & 0.852                & \multicolumn{1}{c|}{\textbf{0.946}} & \textbf{0.865}           \\ \hline
\multirow{6}{*}{R-1000} & \multicolumn{1}{c|}{DeepGini} & \textbf{0.786} & 0.960                & 0.978                & \multicolumn{1}{c|}{0.966}          & \textbf{0.561}       & 0.810                & 0.905                & \multicolumn{1}{c|}{0.893}          & \textbf{0.510}       & 0.830                & 0.914                & \multicolumn{1}{c|}{0.849}          & \textbf{0.847}       & \textbf{0.832}       & 0.661                & \multicolumn{1}{c|}{0.582}          & 0.805                    \\
                        & \multicolumn{1}{c|}{PRIMA}    & 0.744          & 0.942                & 0.994                & \multicolumn{1}{c|}{\textbf{0.998}} & 0.487                & 0.767                & 0.487                & \multicolumn{1}{c|}{0.973}          & 0.452                & 0.690                & 0.919                & \multicolumn{1}{c|}{0.976}          & 0.842                & 0.824                & 0.859                & \multicolumn{1}{c|}{0.933}          & 0.805                    \\
                        & \multicolumn{1}{c|}{MCP}      & 0.691          & 0.874                & 0.821                & \multicolumn{1}{c|}{0.768}          & 0.441                & 0.733                & 0.901                & \multicolumn{1}{c|}{0.862}          & 0.418                & 0.666                & \textbf{0.929}       & \multicolumn{1}{c|}{0.795}          & 0.440                & 0.374                & 0.368                & \multicolumn{1}{c|}{0.348}          & 0.652                    \\
                        & \multicolumn{1}{c|}{DSA}      & 0.436          & 0.897                & 0.870                & \multicolumn{1}{c|}{0.856}          & 0.379                & 0.643                & 0.740                & \multicolumn{1}{c|}{0.694}          & 0.213                & 0.482                & 0.574                & \multicolumn{1}{c|}{0.592}          & 0.644                & 0.516                & 0.694                & \multicolumn{1}{c|}{0.683}          & 0.619                    \\
                        & \multicolumn{1}{c|}{Act}      & 0.428          & 0.964                & 0.933                & \multicolumn{1}{c|}{0.901}          & 0.441                & 0.753                & 0.893                & \multicolumn{1}{c|}{0.904}          & 0.416                & 0.916                & 0.768                & \multicolumn{1}{c|}{0.613}          & 0.813                & 0.797                & 0.664                & \multicolumn{1}{c|}{0.620}          & 0.739                    \\
                        & \multicolumn{1}{c|}{ActGraph} & 0.769          & \textbf{0.971}       & \textbf{0.996}       & \multicolumn{1}{c|}{0.992}          & 0.521                & \textbf{0.857}       & \textbf{0.943}       & \multicolumn{1}{c|}{\textbf{0.984}} & 0.458                & \textbf{0.962}       & 0.916                & \multicolumn{1}{c|}{\textbf{0.977}} & 0.835                & 0.818                & \textbf{0.882}       & \multicolumn{1}{c|}{\textbf{0.952}} & \textbf{0.865}           \\ \hline
\multirow{6}{*}{R-ALL}  & \multicolumn{1}{c|}{DeepGini} & \textbf{0.972} & 0.924                & 0.997                & \multicolumn{1}{c|}{0.997}          & 0.867                & 0.877                & 0.984                & \multicolumn{1}{c|}{0.979}          & \textbf{0.866}       & 0.889                & 0.974                & \multicolumn{1}{c|}{0.976}          & \textbf{0.830}       & 0.849                & 0.865                & \multicolumn{1}{c|}{0.853}          & 0.919                    \\
                        & \multicolumn{1}{c|}{PRIMA}    & 0.963          & 0.920                & 0.999                & \multicolumn{1}{c|}{0.999}          & 0.854                & 0.884                & 0.976                & \multicolumn{1}{c|}{0.992}          & 0.857                & 0.888                & 0.980                & \multicolumn{1}{c|}{0.995}          & 0.824                & 0.857                & \textbf{0.933}       & \multicolumn{1}{c|}{0.981}          & 0.931                    \\
                        & \multicolumn{1}{c|}{MCP}      & 0.955          & 0.842                & 0.873                & \multicolumn{1}{c|}{0.878}          & 0.802                & 0.752                & 0.888                & \multicolumn{1}{c|}{0.859}          & 0.737                & 0.713                & 0.900                & \multicolumn{1}{c|}{0.832}          & 0.515                & 0.508                & 0.491                & \multicolumn{1}{c|}{0.478}          & 0.752                    \\
                        & \multicolumn{1}{c|}{DSA}      & 0.877          & 0.936                & 0.940                & \multicolumn{1}{c|}{0.928}          & 0.796                & 0.832                & 0.936                & \multicolumn{1}{c|}{0.937}          & 0.650                & 0.746                & 0.741                & \multicolumn{1}{c|}{0.737}          & 0.751                & 0.779                & 0.834                & \multicolumn{1}{c|}{0.826}          & 0.828                    \\
                        & \multicolumn{1}{c|}{Act}      & 0.872          & 0.957                & 0.959                & \multicolumn{1}{c|}{0.948}          & 0.476                & 0.874                & 0.962                & \multicolumn{1}{c|}{0.968}          & 0.803                & 0.937                & 0.861                & \multicolumn{1}{c|}{0.819}          & 0.812                & 0.853                & 0.794                & \multicolumn{1}{c|}{0.770}          & 0.854                    \\
                        & \multicolumn{1}{c|}{ActGraph} & 0.962          & \textbf{0.958}       & \textbf{0.999}       & \multicolumn{1}{c|}{\textbf{0.999}} & \textbf{0.869}       & \textbf{0.886}       & \textbf{0.985}       & \multicolumn{1}{c|}{\textbf{0.993}} & 0.858                & \textbf{0.952}       & \textbf{0.980}       & \multicolumn{1}{c|}{\textbf{0.995}} & 0.826                & \textbf{0.861}       & 0.918                & \multicolumn{1}{c|}{\textbf{0.988}} & \textbf{0.939}           \\ \hline
\multicolumn{3}{l}{\textit{R-N denotes RAUC-N.}}                         & \multicolumn{1}{l}{} & \multicolumn{1}{l}{} & \multicolumn{1}{l}{}                & \multicolumn{1}{l}{} & \multicolumn{1}{l}{} & \multicolumn{1}{l}{} & \multicolumn{1}{l}{}                & \multicolumn{1}{l}{} & \multicolumn{1}{l}{} & \multicolumn{1}{l}{} & \multicolumn{1}{l}{}                & \multicolumn{1}{l}{} & \multicolumn{1}{l}{} & \multicolumn{1}{l}{} & \multicolumn{1}{l}{}                & \multicolumn{1}{l}{}    
\end{tabular}
    }
  \label{rq1table}%
\end{table*}%

Specifically, confidence-based methods perform better on FP cases, and embedding-based methods perform better on adversarial cases. In Original, DeepGini performs best with 11 best results. But in C\&W scenario, ActGraph has 13 best results. Especially in \textit{RAUC-100}, ActGraph is 9.41\%$\sim$52.51\% higher than DeepGini.
On the contrary, DSA is an embedding-based method, which performs better in adversarial scenarios than in natural scenarios.
These results also confirm the previous hypothesis that the confidence-based methods work well for natural scenarios and the embedding-based methods work well for adversarial scenarios.

Then, ActGraph outperforms Act, which illustrates that \textit{cnf} is more effective than model activation feature. Because \textit{cnf} not only has the information of neuron activation characteristics, but also the node connection relationship between neurons. In particular, we show that ActGraph also outperforms Act in generalizability in Section~\ref{rq2}.


\textbf{Answer to RQ1}: ActGraph outperforms the baseline methods (i.e., DeepGini, PRIMA, MCP, DSA and Act) in natural and adversarial scenarios.
ActGraph gets 78.13\% best results in the adversarial scenarios and 53.13\% best results in the natural scenarios. In \textit{RAUC-100}, the average results of ActGraph are 5.96\%$\sim$24.36\% higher than the baseline methods.

\subsection{Generalizability of ActGraph}\label{rq2}
In the section, we find the answer to \textbf{RQ2}, validating the performance of ActGraph for prioritization of multiple types of test cases, especially with limited training knowledge. Limited training knowledge means that the trainset of the ranking model contains only one type of cases, which can trigger DNN vulnerabilities, but multiple types of test cases need to be prioritized in the testing phase.

\textbf{Implementation details.}
(1) Five training types are set up to explore the generalizability of ActGraph to \textbf{Mix} testset.
(2) The size of trainset of ranking model is 2,000, which contains 1,000 positive cases and 1,000 negative cases.
(3) \textit{RAUC-100} and \textit{RAUC-500} are evaluated in the experiment, since the time and cost of prioritization is limited.

\textbf{Results and analysis.}
The results are shown in Table~\ref{rq2table}.
Since DeepGini and MCP are unsupervised methods, their prioritization results are not affected by the type of testset.
In the total 40 results, ActGraph shows 29 best results (72.5\%), followed by PRIMA with 8 best results (20\%), MCP with 2 best results and DeepGini with zero best results.
Then, we average the 4 metrics, and ActGraph performs the best, followed by PRIMA.
The average results of ActGraph are 0.878$\sim$0.898, which are 3.58\%$\sim$4.65\% higher than PRIMA, 9.02\%$\sim$11.08\% higher than DeepGini, 6.46\%$\sim$6.49\% higher than Act, 22.70\%$\sim$23.54\% higher than DSA, and 10.90\%$\sim$16.09\% higher than MCP.
Especially, ActGraph performs better in \textit{RAUC-100}, is 4.65\%$\sim$22.70\% higher than the other baseline methods.

\begin{table}[t]
  \centering
  \caption{The results of prioritization in mixed test case scenarios.}
  \resizebox{1\linewidth}{!}{
    \Huge
\begin{tabular}{ccccccccccccccccccccccc}
\hline
\multicolumn{2}{c|}{\textbf{Datasets}}                 & \multicolumn{5}{c|}{MNIST}                                                                                          & \multicolumn{10}{c|}{CIFAR-10}                                                                                                                                                                                                                                    & \multicolumn{5}{c|}{CIFAR-100}                                                                                                  & \multirow{3}{*}{Average} \\ \cline{1-22}
\multicolumn{2}{c|}{\textbf{Models}}                   & \multicolumn{5}{c|}{LeNet-5}                                                                                        & \multicolumn{5}{c|}{VGG16}                                                                                                      & \multicolumn{5}{c|}{ResNet18}                                                                                                   & \multicolumn{5}{c|}{VGG19}                                                                                                      &                          \\ \cline{1-22}
\multicolumn{2}{c|}{\textbf{Train type}}               & Original       & Rotate         & JSMA                 & CW                   & \multicolumn{1}{c|}{Mix}            & Original             & Rotate               & JSMA                 & CW                   & \multicolumn{1}{c|}{Mix}            & Original             & Rotate               & JSMA                 & CW                   & \multicolumn{1}{c|}{Mix}            & Original             & Rotate               & JSMA                 & CW                   & \multicolumn{1}{c|}{Mix}            &                          \\ \hline
\multirow{6}{*}{R-100} & \multicolumn{1}{c|}{DeepGini} & 0.899          & 0.899          & 0.899                & 0.899                & \multicolumn{1}{c|}{0.899}          & 0.738                & 0.738                & 0.738                & 0.738                & \multicolumn{1}{c|}{0.738}          & 0.828                & 0.828                & 0.828                & 0.828                & \multicolumn{1}{c|}{0.828}          & 0.682                & 0.682                & 0.682                & 0.682                & \multicolumn{1}{c|}{0.682}          & 0.787                    \\
                       & \multicolumn{1}{c|}{PRIMA}    & 0.944          & 0.933          & 0.984                & 0.997                & \multicolumn{1}{c|}{0.957}          & 0.750                & 0.745                & 0.780                & 0.912                & \multicolumn{1}{c|}{0.890}          & 0.736                & 0.751                & 0.839                & 0.968                & \multicolumn{1}{c|}{\textbf{0.897}} & 0.807                & 0.776                & \textbf{0.848}       & \textbf{0.852}       & \multicolumn{1}{c|}{0.659}          & 0.851                    \\
                       & \multicolumn{1}{c|}{MCP}      & 0.964          & 0.964          & 0.964                & 0.964                & \multicolumn{1}{c|}{0.964}          & 0.858                & \textbf{0.858}       & 0.858                & 0.858                & \multicolumn{1}{c|}{0.858}          & 0.853                & \textbf{0.853}       & 0.853                & 0.853                & \multicolumn{1}{c|}{0.853}          & 0.479                & 0.479                & 0.479                & 0.479                & \multicolumn{1}{c|}{0.479}          & 0.789                    \\
                       & \multicolumn{1}{c|}{DSA}      & 0.930          & 0.772          & 0.789                & 0.925                & \multicolumn{1}{c|}{0.833}          & 0.728                & 0.508                & 0.511                & 0.662                & \multicolumn{1}{c|}{0.475}          & 0.376                & 0.665                & 0.629                & 0.432                & \multicolumn{1}{c|}{0.457}          & 0.541                & 0.761                & 0.781                & 0.837                & \multicolumn{1}{c|}{0.802}          & 0.671                    \\
                       & \multicolumn{1}{c|}{Act}      & 0.892          & 0.856          & 0.960                & 0.940                & \multicolumn{1}{c|}{0.899}          & 0.864                & 0.785                & 0.864                & \textbf{0.944}       & \multicolumn{1}{c|}{0.900}          & 0.770                & 0.568                & 0.854                & 0.921                & \multicolumn{1}{c|}{0.841}          & 0.751                & 0.742                & 0.764                & 0.747                & \multicolumn{1}{c|}{0.796}          & 0.833                    \\
                       & \multicolumn{1}{c|}{ActGraph} & \textbf{0.967} & \textbf{0.979} & \textbf{1.000}       & \textbf{1.000}       & \multicolumn{1}{c|}{\textbf{0.985}} & \textbf{1.000}       & 0.838                & \textbf{0.905}       & 0.898                & \multicolumn{1}{c|}{\textbf{0.970}} & \textbf{0.867}       & 0.853                & \textbf{0.857}       & \textbf{0.978}       & \multicolumn{1}{c|}{0.861}          & \textbf{0.849}       & \textbf{0.837}       & 0.765                & 0.727                & \multicolumn{1}{c|}{\textbf{0.818}} & \textbf{0.898}           \\ \hline
\multirow{6}{*}{R-500} & \multicolumn{1}{c|}{DeepGini} & 0.908          & 0.908          & 0.908                & 0.908                & \multicolumn{1}{c|}{0.908}          & 0.798                & 0.798                & 0.798                & 0.798                & \multicolumn{1}{c|}{0.798}          & 0.766                & 0.766                & 0.766                & 0.766                & \multicolumn{1}{c|}{0.766}          & 0.678                & 0.678                & 0.678                & 0.678                & \multicolumn{1}{c|}{0.678}          & 0.788                    \\
                       & \multicolumn{1}{c|}{PRIMA}    & 0.903          & 0.906          & 0.979                & 0.991                & \multicolumn{1}{c|}{0.929}          & 0.740                & 0.726                & 0.794                & 0.910                & \multicolumn{1}{c|}{0.886}          & 0.744                & 0.721                & 0.831                & \textbf{0.968}       & \multicolumn{1}{c|}{\textbf{0.887}} & \textbf{0.758}       & 0.779                & \textbf{0.792}       & \textbf{0.847}       & \multicolumn{1}{c|}{0.748}          & 0.842                    \\
                       & \multicolumn{1}{c|}{MCP}      & 0.909          & 0.909          & 0.909                & 0.909                & \multicolumn{1}{c|}{0.909}          & 0.773                & 0.773                & 0.773                & 0.773                & \multicolumn{1}{c|}{0.773}          & 0.732                & 0.732                & 0.732                & 0.732                & \multicolumn{1}{c|}{0.732}          & 0.453                & 0.453                & 0.453                & 0.453                & \multicolumn{1}{c|}{0.453}          & 0.717                    \\
                       & \multicolumn{1}{c|}{DSA}      & 0.850          & 0.799          & 0.807                & 0.839                & \multicolumn{1}{c|}{0.825}          & 0.618                & 0.508                & 0.590                & 0.666                & \multicolumn{1}{c|}{0.532}          & 0.479                & 0.538                & 0.515                & 0.455                & \multicolumn{1}{c|}{0.424}          & 0.615                & 0.678                & 0.701                & 0.689                & \multicolumn{1}{c|}{0.722}          & 0.642                    \\
                       & \multicolumn{1}{c|}{Act}      & 0.905          & 0.878          & 0.963                & 0.958                & \multicolumn{1}{c|}{0.857}          & 0.856                & 0.717                & 0.878                & 0.935                & \multicolumn{1}{c|}{0.889}          & 0.706                & 0.670                & 0.756                & 0.785                & \multicolumn{1}{c|}{0.793}          & 0.750                & 0.765                & 0.737                & 0.724                & \multicolumn{1}{c|}{0.742}          & 0.813                    \\
                       & \multicolumn{1}{c|}{ActGraph} & \textbf{0.965} & \textbf{0.957} & \textbf{0.986}       & \textbf{0.991}       & \multicolumn{1}{c|}{\textbf{0.971}} & \textbf{0.984}       & \textbf{0.869}       & \textbf{0.912}       & \textbf{0.944}       & \multicolumn{1}{c|}{\textbf{0.946}} & \textbf{0.786}       & \textbf{0.783}       & \textbf{0.862}       & 0.967                & \multicolumn{1}{c|}{0.852}          & 0.731                & \textbf{0.833}       & 0.720                & 0.728                & \multicolumn{1}{c|}{\textbf{0.767}} & \textbf{0.878}           \\ \hline
\multicolumn{4}{l}{\textit{$\bullet$ R-N denotes RAUC-N.}}                               & \multicolumn{1}{l}{} & \multicolumn{1}{l}{} & \multicolumn{1}{l}{}                & \multicolumn{1}{l}{} & \multicolumn{1}{l}{} & \multicolumn{1}{l}{} & \multicolumn{1}{l}{} & \multicolumn{1}{l}{}                & \multicolumn{1}{l}{} & \multicolumn{1}{l}{} & \multicolumn{1}{l}{} & \multicolumn{1}{l}{} & \multicolumn{1}{l}{}                & \multicolumn{1}{l}{} & \multicolumn{1}{l}{} & \multicolumn{1}{l}{} & \multicolumn{1}{l}{} & \multicolumn{1}{l}{}                & \multicolumn{1}{l}{}    
\end{tabular}
    }
  \label{rq2table}%
\end{table}%


In the \textit{RAUC-100} of VGG16, ResNet18 and VGG19, the variance of PRIMA from 0.0062 to 0.0095, DSA from 0.0123 to 0.0163, Act from 0.0005 to 0.0184. The variance of ActGraph achieve 0.0041, 0.0028 and 0.0027, respectively.
As the result, ActGraph performs more consistently than other supervised learning baseline methods. This shows the stable effectiveness of ActGraph for different types of test cases in limited training knowledge.
This also shows that, for the DNN model under test, the ranking model of ActGraph does not need to be retrained frequently, because it can perform stably on different types of test cases, which indicates the generalizability of ActGraph.

\textbf{Answer to RQ2}: ActGraph outperforms the baseline methods (i.e., DeepGini, PRIMA, MCP, DSA and Act) in mixed scenarios.
ActGraph has 72.5\% of the best results, especially in average \textit{RAUC-100}, which is 4.65\%$\sim$22.70\% higher than the baseline methods.
In addition, ActGraph shows better stability than the baseline methods by calculating the variance of \textit{RAUC-100}. The variance of the baseline methods are 3.39 to 6.57 times that of ActGraph.

\subsection{Interpretability of ActGraph}\label{rq3}
In the section, we find the answer to \textbf{RQ3}, explaining why ActGraph can be used effectively for prioritization. We show the visualization of test cases with high prioritization, and carry out qualitative analysis and quantitative analysis.

\textbf{Implementation details.}
(1) The t-SNE visualization for qualitative analysis and the heat map for quantitative analysis.
(2) We use mixed cases to analyze the intra-class and inter-class distances of the features of ActGraph and baseline algorithms.

\begin{figure}[htbp]
  \centering
  \subfigure
  {
  \includegraphics[width=0.8\textwidth]{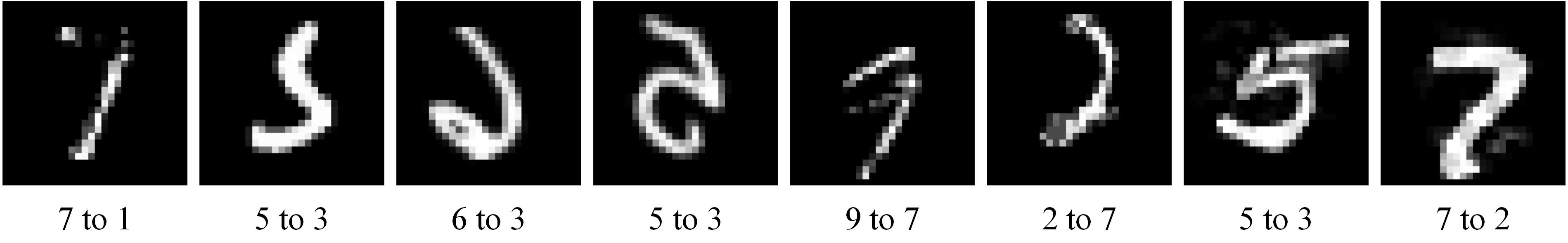}
    \label{select_mnist}
    }
  \subfigure
  {
  \includegraphics[width=0.8\textwidth]{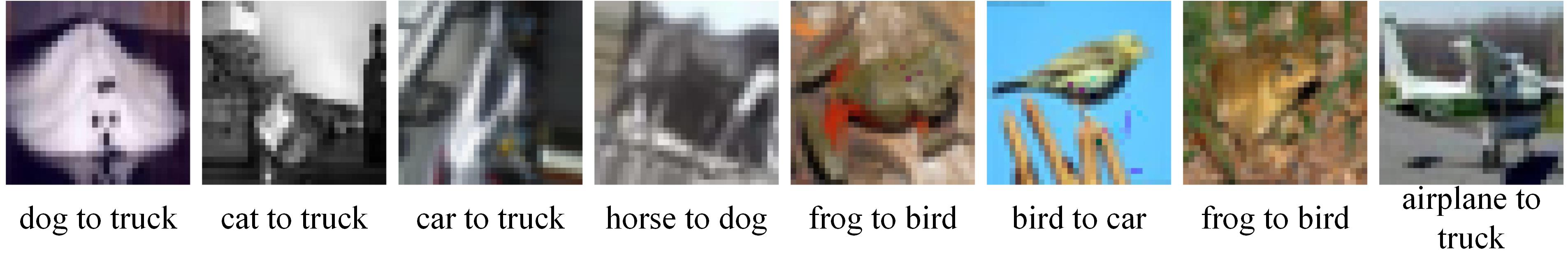}
    \label{select_cifar10}
    }
  \subfigure
  {
  \includegraphics[width=0.8\textwidth]{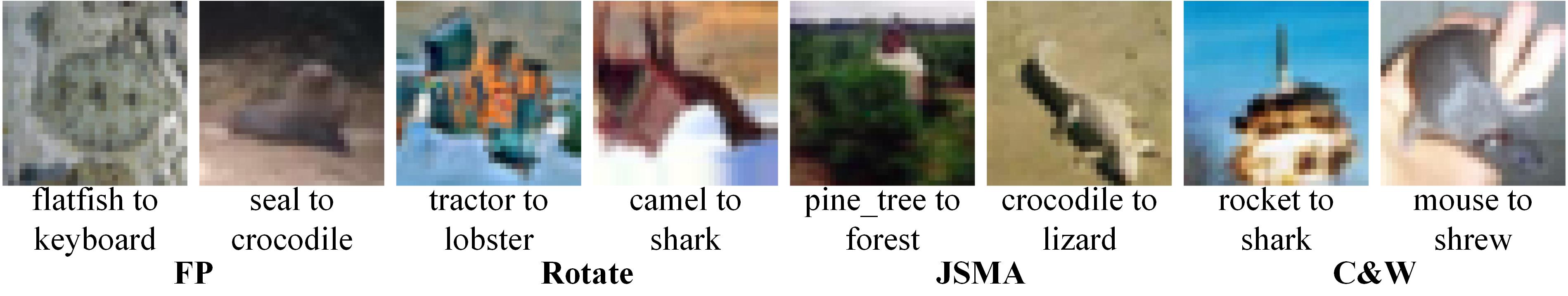}
    \label{select_cifar100}
    }
  \caption{The visualization of test cases with high prioritization (i.e. FP, Rotate, JSMA and C\&W). The first row is from MNIST, second row is from CIFAR-10 and third row is from CIFAR-100.
  }
  \label{select_sample}
\end{figure}

\textbf{Test cases visualization.}
The visualization of test cases with high prioritization is shown in Fig.~\ref{select_sample}. Intuitively, for the FP and Rotate test cases, they are also difficult to recognize by humans. For example, the images ``7" and ``5" are incomplete; The image ``dog" is too bright, and the hair is too long, which blocks the basic features of dog; The colors of the images ``cat", ``flatfish", ``seal" and ``horse" are similar to the environment. In particular, the ``tractor" and ``camel" are rotated 180 degrees resulting in the blue sky at the bottom of the image, thus identifying them as ``lobster" and ``shark". For JSMA and C\&W, some images are also broken or blurred, such as ``9'' and ``2'' in the first line and ``pine tree'' in the third line. Most of the images are clear, but adding unobserved adversarial perturbations causes DNN output errors. This shows that ActGraph can pick up weak adversarial perturbations.

\begin{figure}[htbp]
  \centering
  \subfigure[Confidence]{
  \includegraphics[width=0.21\textwidth]{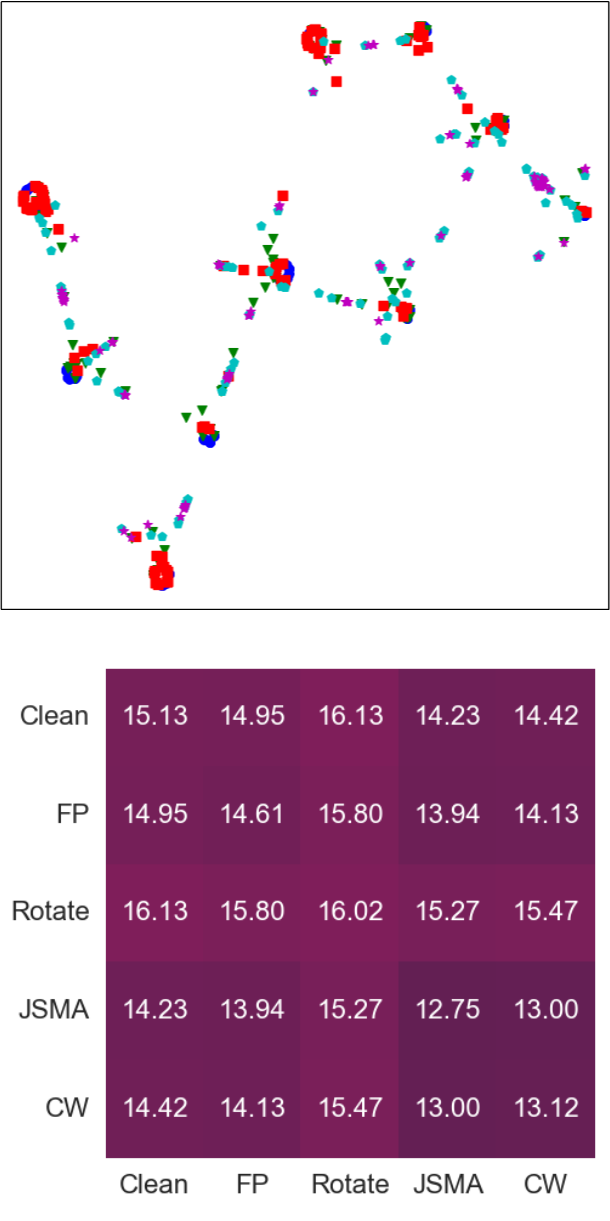}
    \label{confidence_tsne}
    }
  \subfigure[Embedding]
  {
  \includegraphics[width=0.21\textwidth]{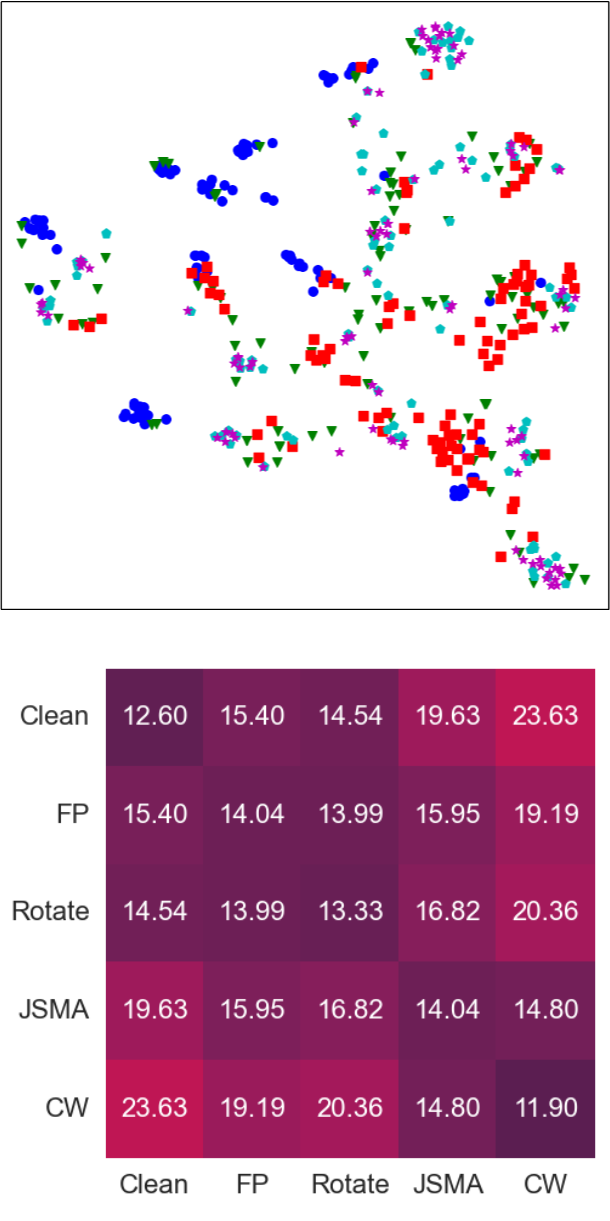}
    \label{embed_tsne}
    }
  \subfigure[PRIMA]
  {
  \includegraphics[width=0.21\textwidth]{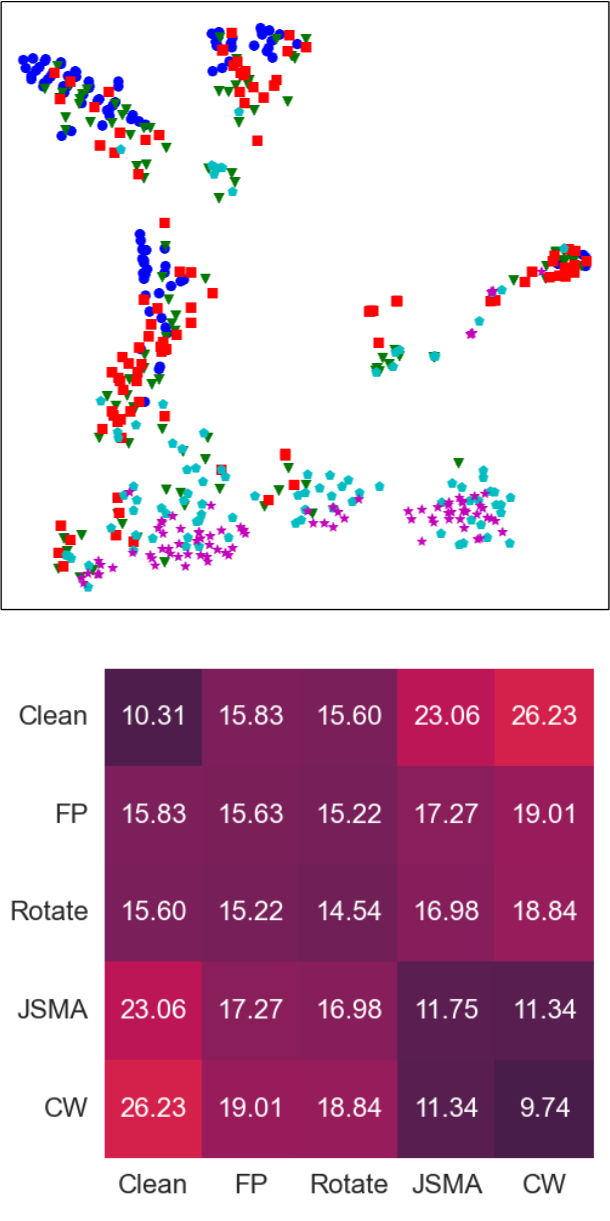}
    \label{prima_tsne}
    }
  \subfigure[ActGraph]
  {
  \includegraphics[width=0.244\textwidth]{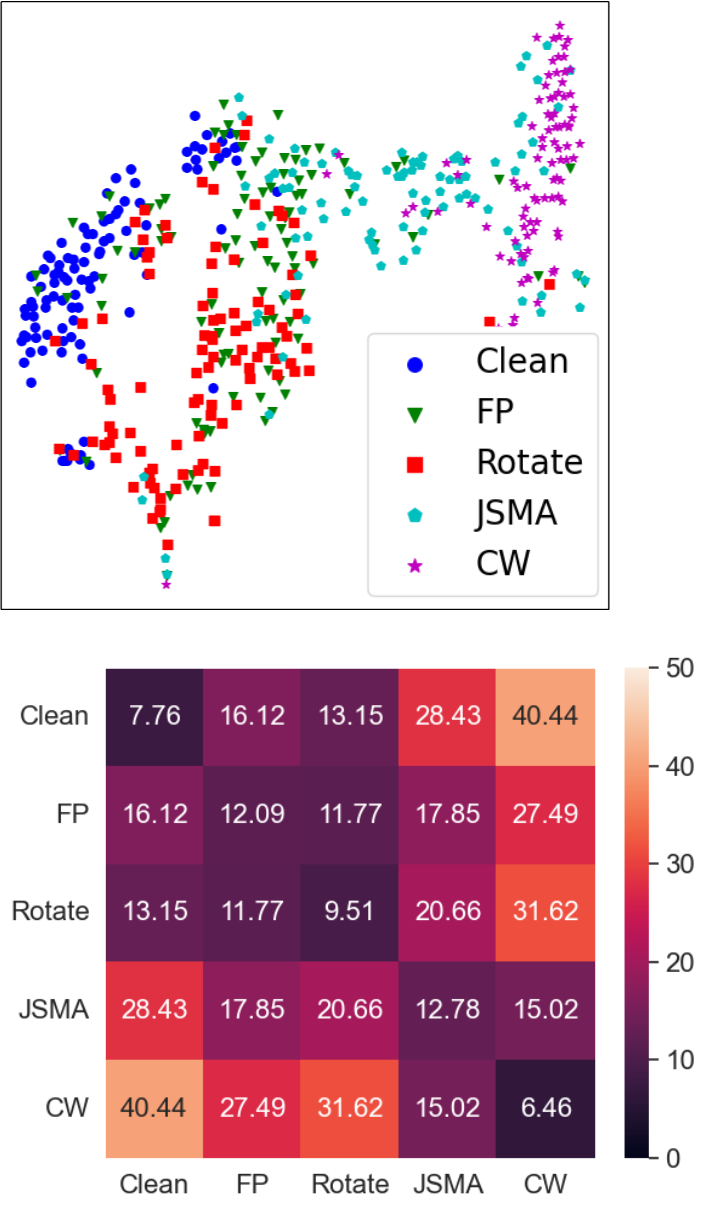}
    \label{graph_tsne}
    }
  \caption{The t-SNE and heatmap visualization. (a) Confidence is the output from the DNN confidence layer. (b) Embedding is the output from the last hidden layer of the DNN. (c) PRIMA represents the mutation feature from the PRIMA method. (d) ActGraph represents the center node feature. The number of cases of each type is 200. It is a VGG16 trained on CIFAR-10.
  }
  \label{tsne}
\end{figure}

\textbf{Qualitative analysis.}
The visualization of t-SNE is the first row of Fig.~\ref{tsne} for qualitative analysis. Intuitively, the features of Confidence and Embedding are confused, and the distances between different types are relatively close.
In PRIMA, Clean and FP are close in distance, JSMA and C\&W are close in distance, and Rotate is in the middle between natural and adversarial cases. This shows that although PRIMA can distinguish between FP and adversarial cases, it is difficult to distinguish between Clean and FP cases.
In ActGraph, Rotate and Clean are distinguished, and most of the FP and Rotate overlap, only a few FP and Clean intersect, and the distance between JSMA and C\&W is also farther than PRIMA.
This shows that the center node feature of ActGraph not only has better prioritization performance for adversarial cases, but also has better prioritization effect on natural cases and FP cases than existing methods.

\textbf{Quantitative analysis.}
We calculate intra-class and inter-class distances on the t-SNE for five types of cases, represented by the heatmap, for quantitative analysis, where the distance measure used Euclidean distance. The heat map is the second row of Fig.~\ref{tsne}.
For intra-class distance, ActGraph is 6.46$\sim$12.78. Except JSMA, the intra-class distance of ActGraph is smaller than other methods.
For inter-class distance, intuitively, the distance between natural cases (Clean, FP and Rotate) and adversarial cases (JSMA and C\&W) of ActGraph is all the farthest, with the farthest distance being 40.44, which is 1.54$\sim$2.80 times of the other three methods. This shows that ActGraph can distinguish natural samples from adversarial samples better than other methods.
In addition, the distance from FP to Clean of ActGraph is 16.12, the distance from FP to Rotate is 11.77, and the difference is 4.35, while the difference of Confidence is -0.85, Embedding is 1.41, and PRIMA is 0.61. This shows that the FP of ActGraph is closer to Rotate and farther from Clean, and the effect of prioritization will be better. Also, ActGraph's JSMA to C\&W distance is farther than other methods.

\textbf{Answer to RQ3}: ActGraph has smaller intra-class distances and larger inter-class distances than the baseline methods. For inter-class distances, the maximum inter-class distance of ActGraph is 1.54$\sim$2.80 times that of the baseline methods, and the average inter-class distance of ActGraph is 1.12$\sim$1.36 times that of the baseline methods. For inter-class distances, the minimum inter-class distance of the baseline methods are 1.52$\sim$2.04 times that of ActGraph and the average inter-class distance are 1.28$\sim$1.47 times that of ActGraph.

\subsection{Sensitivity Analysis of Parameters}
\label{rq4}
In this section, we find the answer to \textbf{RQ4}. We analyze the influence of the size of trainset and training parameters of ActGraph.

\textbf{Implementation details.}
(1) We set \textit{RAUC-100} as the evaluation metric.
(2) It is a VGG16 trained on CIFAR-10.
(3) The type of trainset and testset are both Mix.

\begin{figure}[t]
  \centering
  \subfigure[LeNet-5]{
  \includegraphics[width=0.4\textwidth]{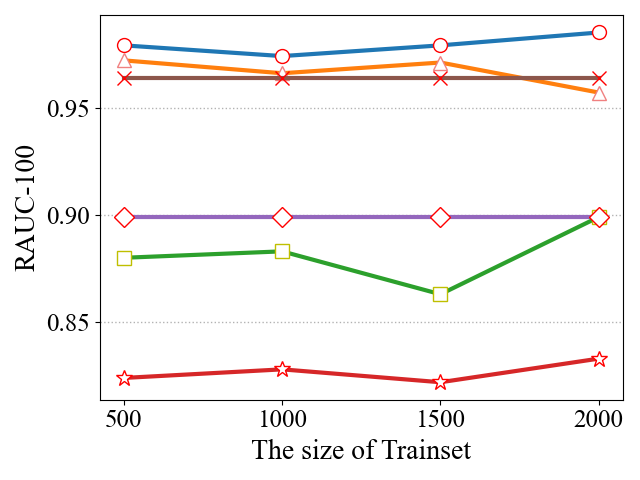}
    \label{train_num_mnist}
    }
  \subfigure[VGG16]
  {
  \includegraphics[width=0.4\textwidth]{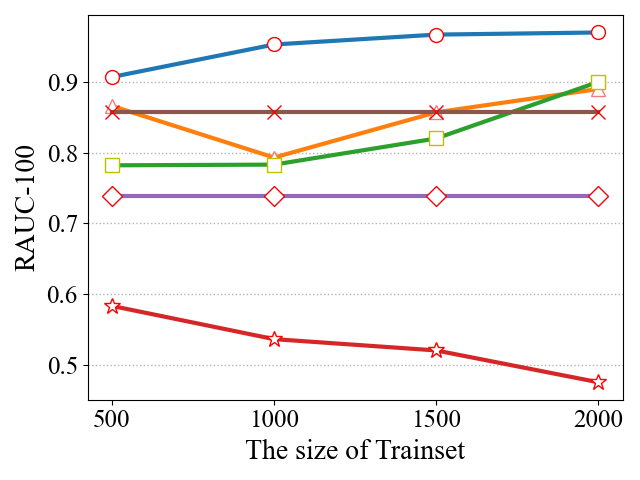}
    \label{train_num_cifar10}
    }
  \subfigure[ResNet18]
  {
  \includegraphics[width=0.4\textwidth]{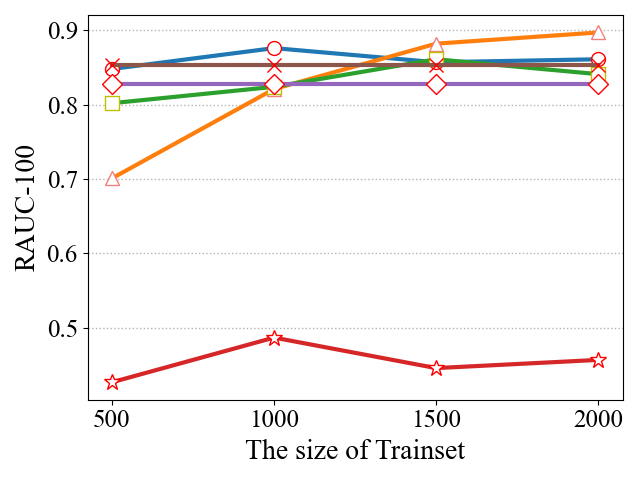}
    \label{train_num_cifar10_res}
    }
  \subfigure[VGG19]
  {
  \includegraphics[width=0.4\textwidth]{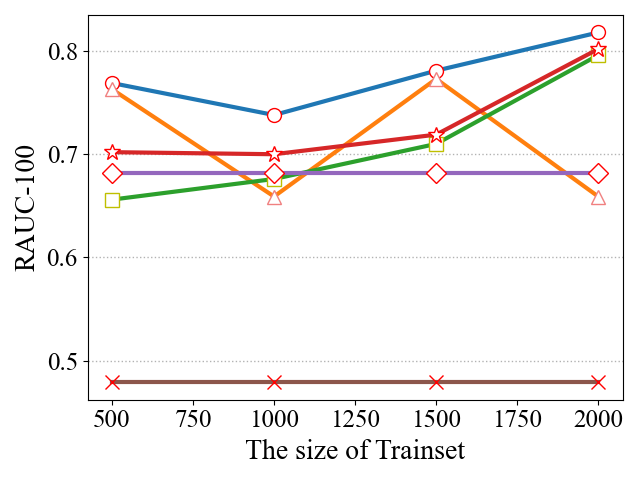}
    \label{train_num_cifar100}
    }
  \subfigure
  {
  \includegraphics[width=0.7\textwidth]{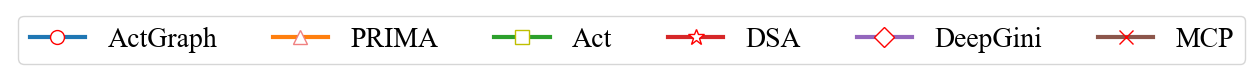}
  \label{legend}
  }
  \caption{Influence of different size of trainset.}
  \label{train_num}
\end{figure}

\textbf{Influence of trainset size.}
The result shows in Fig.~\ref{train_num}. DeepGini and MCP are unsupervised methods, so are not affected by the size of the trainset. In LeNet-5 and VGG (VGG16 and VGG19), ActGraph exhibits the best prioritization performance, and the performance increases with the size of the dataset. In ResNet18, ActGraph performs better than most baseline algorithms. Importantly, ActGraph is less affected by changes in trainset size, while the remaining three baseline algorithms (i.e. PRIMA, DSA and Act) are more affected by changes in trainset size. This shows that ActGraph can learn effective features from a small number of cases.

\begin{figure}[htbp]
  \centering
  \subfigure{
  \includegraphics[width=0.31\textwidth]{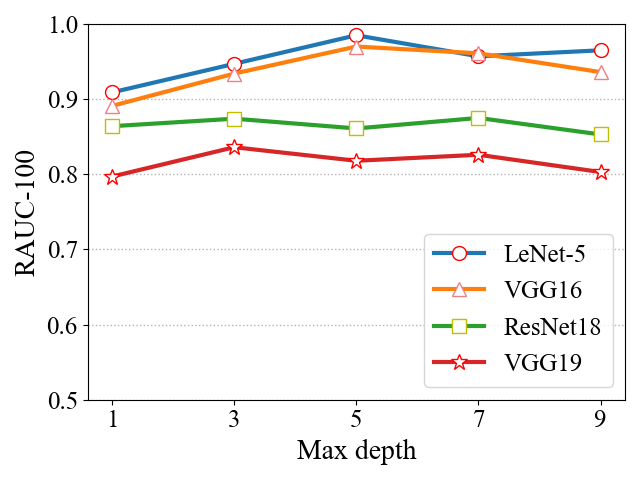}
    \label{maxdepth}
    }
  \subfigure
  {
  \includegraphics[width=0.31\textwidth]{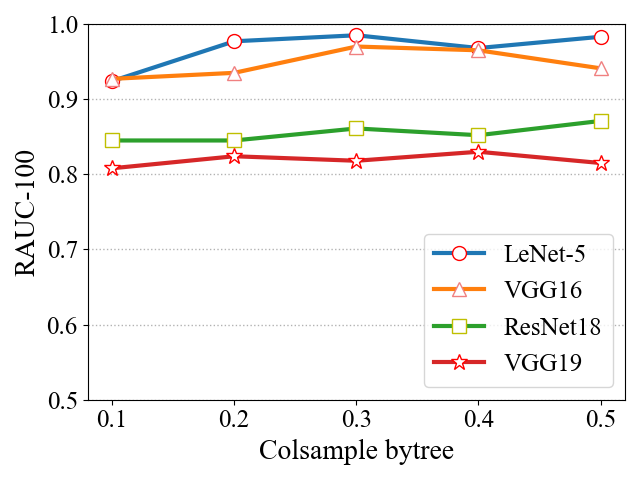}
    \label{sampletree}
    }
  \subfigure
  {
  \includegraphics[width=0.31\textwidth]{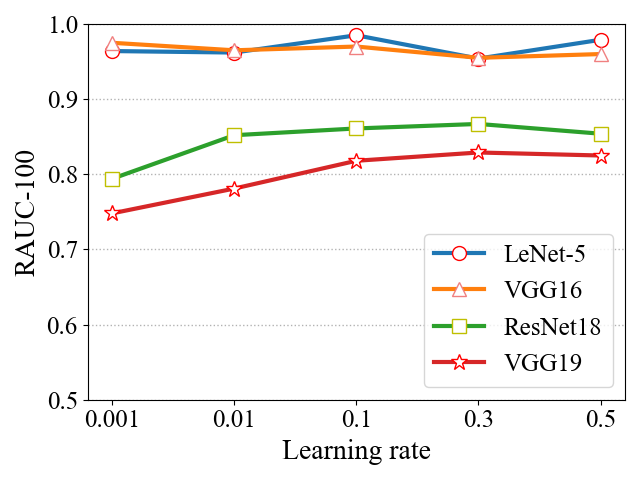}
    \label{learningrate}
    }
  \caption{Influence of hyperparameters in ActGraph.}
  \label{train_para}
\end{figure}

\textbf{Influence of training parameters.} We investigate the influence of main parameters in ActGraph, including \textit{Max depth} (the maximum tree depth for each XGBoost model), \textit{Colsample bytree} (the sampling ratio of columns of features when constructing each tree) and \textit{Learning rate} (the boosting learning rate) in the XGBoost ranking algorithm. Fig.~\ref{train_para} shows the effectiveness of ActGraph under different parameter settings in \textit{RAUC-100} across the four models. We found that, ActGraph performs stably under different parameter settings and our default settings are proper.

\textbf{Answer to RQ4}: For the trainset size, we set four scenarios ranging from 500 to 2,000. ActGraph has stable performance and increases with the increase of the trainset size. For training hyperparameters, ActGraph is also stable. These results indicate that our previous parameter settings are appropriate.

\subsection{Time Complexity}\label{rq5}
In this section, we find the answer to \textbf{RQ5}, referring to the prioritization time cost.

\textbf{Implementation details.}
We measure average running time for ranking 10,000 test cases by ActGraph and baselines. We run each method for 5 times, and the average is identified as the final result.

\textbf{Results and analysis.} Firstly, we theoretically analyze the complexity of ActGraph according to different steps.

The time complexity of ActGraph includes acquiring activation values of multi-layer neurons, calculating $ap$, calculating $nf$ and calculating \textit{cnf}, so its time complexity is
\begin{equation}\label{time complexity}
    T\sim O(t\times V)+  O(t\times E)+  O(t\times E)+  O(t\times E)\sim O(t\times E)
\end{equation}
where $t$ is the number of samples, $V$ is the number of neurons, and $E$ is the number of edges between neurons.

Further, we analyze the efficiency of ActGraph from the real running time. According to the Table~\ref{rq5table}, the running time of ActGraph is acceptable, the time cost of ActGraph increases due to the increase of the total number of neurons and edges. In general, ActGraph is much faster than PRIMA, but is inferior to other methods. The reason is that we search more layers and neurons. Besides, ActGraph calculates the weighted edges between neurons, and calculates the center node features in the activation graph. PRIMA runs on average 50 times longer than ActGraph.

\begin{table}[htbp]
  \centering
  \caption{Time (seconds) taken to prioritize 10,000 test cases}
  \resizebox{0.6\linewidth}{!}{
    \begin{tabular}{cccccccc}
    \hline
    \multirow{2}{*}{\textbf{Datasets}} & \multirow{2}{*}{\textbf{Models}} & \multicolumn{6}{c}{\textbf{Methods}}                   \\
                                       &                                  & DeepGini & PRIMA    & Act   & MCP   & DSA   & ActGraph \\ \hline
    MNIST                              & LeNet-5                          & 0.662    & 8049.53  & 0.642 & 0.626 & 11.59 & 103.69   \\ \hline
    \multirow{2}{*}{CIFAR-10}          & VGG16                            & 1.274    & 10864.19 & 1.115 & 1.465 & 19.09 & 246.26   \\
                                       & ResNet18                         & 1.843    & 14516.53 & 1.735 & 2.444 & 37.36 & 306.44   \\ \hline
    CIFAR-100                          & VGG19                            & 1.380    & 15461.26 & 1.367 & 1.705 & 19.00 & 291.05   \\ \hline
    \end{tabular}
  }
  \label{rq5table}%
\end{table}%

\textbf{Answer to RQ5}: The average running time for ActGraph to prioritize 10,000 cases is about four minutes, which is acceptable.

\section{THREATS TO VALIDITY}
\label{threat}
\textbf{The depth of the graph}.
We explain why $K$=4 and take only two layers of \textit{cnf} in Section~\ref{proof}.
Too large $K$ value (more than 4) and the layer number of \textit{cnf} (more than 2) will not only increase the time cost, but also introduce noise from the shallow layer.
And the value of $K$ too small to obtain the valid \textit{cnf} value.
Therefore, it is appropriate to set $K$=4, which is also confirmed by the comprehensive experimental results.

\textbf{Time cost}.
The time cost of ActGraph may be affected by the neuron number of the DNN. We demonstrate that the ActGraph's runtime is acceptable by applying three popular DNN models with significant differences in the number of neurons.

\textbf{Access to DNNs}.
ActGraph is a white-box method of prioritizing test cases.
It needs to access the DNN for model weight parameters and deep activation of test cases.
It is widely accepted that DNN testing could have full knowledge of the target model in software engineering.




\section{CONCLUSIONS\label{conclusions}}
Aiming at the problems of limited application scenarios and high time cost of existing test case prioritization methods, we propose a test case prioritization method based on the DNN activation graph, named ActGraph.
We observe that the activation graphs of cases that trigger model vulnerabilities different from those of normal cases significantly.
Motivated by it, ActGraph extracts the node features and adjacency matrix of test cases by building an activation graph, and uses the message passing mechanism to aggregate node features and adjacency matrix to obtain more effective center node features for test case prioritization.
Extensive experiments have verified the effectiveness of ActGraph, which outperforms the SOTA method in both natural and adversarial scenarios, especially in \textit{RAUC-100} ($\sim\times$1.40).
And when the number of test cases is 10,000, the actual running time of the SOTA method is 50 times that of ActGraph.
The experiments show that ActGraph has significantly better performance in terms of effectiveness, generalizability, and efficiency.

In the future, we will improve our ActGraph approach and apply it to more popular tasks and models, such as the transformer model for natural language processing and the long short-term memory model for time series forecasting.

\section*{ACKNOWLEDGMENT}
This research was supported by
the National Natural Science Foundation of China, (No. 62072406) 
the National Key Laboratory of Science and Technology on Information System Security (No. 61421110502), 
the National Natural Science Foundation of China (No. 62103374). 


\end{document}